\documentclass{aa}
\usepackage{graphicx,verbatim}
\usepackage{graphicx}
\usepackage{booktabs}
\usepackage{longtable}

\usepackage{txfonts}
\pdfoutput=1

\def\simlt{\lower.5ex\hbox{$\; \buildrel < \over \sim \;$}}
\def\simgt{\lower.5ex\hbox{$\; \buildrel > \over \sim \;$}}
\def\ltsim{\raise 2pt \hbox {$<$} \kern-1.1em \lower 4pt \hbox {$\sim$}}
\def\ltapprox{\raise 2pt \hbox {$<$} \kern-1.1em \lower 5pt \hbox {$\approx
$}}
\def\gtsim{\raise 2pt \hbox {$>$} \kern-1.1em \lower 4pt \hbox {$\sim$}}
\def\gtapprox{\raise 2pt \hbox {$>$} \kern-1.1em \lower 5pt \hbox {$\approx
$}}

\begin{document}

\title{The two-component giant radio halo in the galaxy cluster Abell\,2142} 
\author{T.~Venturi\inst{1}, M.~Rossetti\inst{2}, G.~Brunetti\inst{1},
  D.~Farnsworth\inst{3,4}, F.~Gastaldello\inst{2}, S.~Giacintucci\inst{5,6},
  D.V.~Lal\inst{7},  L.~Rudnick\inst{3}, T.W.~Shimwell\inst{8},
  D.~Eckert\inst{9}, S.~Molendi\inst{2}, M.~Owers\inst{10,11} 
}
\institute
{
INAF -- Istituto di Radioastronomia, via Gobetti 101, I-40129, Bologna, Italy\\
\email{tventuri@ira.inaf.it}
\and
INAF -- IASF-Milano, Via Bassini 15, 20133 Milano, Italy
\and
Minnesota Institute for Astrophysics, School of Physics and Astronomy, University of
Minnesota, 116 Church Street SE, Minneapolis, MN, 55455, USA
\and
Cray, Inc., 380 Jackson Street, Suite 210, St. Paul, MN 55101, USA
\and
Naval Research Laboratory, Washington, DC 20375, USA
\and
Department of Astronomy, University of Maryland, College Park, MD, 20742-2421, USA
\and
National Centre for Radio Astrophysics, T. I. F. R., Post Bag 3, Ganeshkhind, Pune 411007
\and
Leiden Observatory, Leiden University, PO Box 9513, NL-2300 RA Leiden, the Netherlands
\and
Department of Astronomy, University of Geneva, ch. d'Ecogia 16, 1290 Versoix, Switzerland
\and
Australian Astronomical Observatory, PO Box 915, North Ryde, NSW 1670, Australia
\and
Department of Physics and Astronomy, Maquarie University, NSW 2109, Australia
}

\date{Received 00 - 00 - 0000; accepted 00 - 00 - 0000}

\titlerunning{The two--component giant radio halo in A\,2142}
\authorrunning{Venturi et al.}

\abstract
{}
{We report on a spectral study at radio frequencies of the 
  giant radio halo in A\,2142 (z=0.0909), which we performed to explore
  its nature and origin.
  The optical and X-ray properties of the cluster suggest that A\,2142 is not a major
  merger and the presence of a giant radio halo is somewhat surprising.}
{We performed deep radio observations of A\,2142 with the Giant Metrewave
  Radio Telescope (GMRT) at 608 MHz, 
322 MHz, and 234 MHz and with the Very Large Array (VLA) in the 1--2 GHz band.
We obtained high-quality images at all frequencies in a wide
range of resolutions, from the galaxy scale, i.e. $\sim5^{\prime\prime}$, 
up to $\sim 60^{\prime\prime}$ to image the diffuse cluster--scale emission.
The radio halo is well detected at all frequencies and extends out to 
the most distant cold front in A\,2142, about 1 Mpc away from the 
cluster centre. We studied the spectral index in two regions: the central 
part of the halo, where the X--ray emission peaks and the two brightest 
dominant galaxies are located; and a second region, known as the ridge
(in the direction 
of the most distant south--eastern cold front), selected to follow the bright 
part of the halo and X-ray emission.
We complemented our deep 
observations with a preliminary LOw Frequency ARray (LOFAR) image at 118 MHz
and with the re-analysis of archival VLA data at 1.4 GHz.} 
{The two components of the radio halo show different observational properties.
The central brightest part has higher surface brightess and a spectrum whose
steepness is similar to those of the known radio halos, i.e. 
$\alpha^{\rm 1.78~GHz}_{\rm 118~MHz}=1.33\pm 0.08$.
The ridge, which fades into the larger scale emission, is broader in size and has
considerably lower surface brightess and a moderately steeper spectrum,
i.e. $\alpha^{\rm 1.78~GHz}_{\rm 118~MHz}\sim 1.5$.
We propose that the brightest part of
the radio halo is powered by the central sloshing in A\,2142, in a process
similar to what has been suggested for mini-halos, or by secondary electrons
generated by hadronic collisions in the ICM.
On the other hand, the steeper ridge may probe particle re-acceleration
by turbulence generated either by stirring the gas and magnetic fields 
on a larger scale or by less energetic mechanisms, such as continuous
infall of galaxy groups or an off-axis (minor) merger.}
{}

\maketitle
\section{Introduction}\label{sec:intro}

Cluster mergers are the most energetic events in the Universe. With total 
energy outputs of order $10^{63}- 10^{64}$ erg,
mergers are a natural way to account for mass assembly: galaxy clusters 
form as a consequence of merger trees to reach and exceed masses of order 
10$^{15}$ M$_{\odot}$. 
The gravitational energy released into the cluster volume during mergers 
deeply affects the dynamics of the galaxies and the properties of the 
thermal intracluster medium (ICM) and non-thermal relativistic particle 
and magnetic field emission (Brunetti \& Jones \cite{bj14}).
The impressive quality of the X--ray  $Chandra$ and {\it XMM--Newton} images
(e.g. Markevitch et al. \cite{markevitch00}, Rossetti et al. \cite{rossetti13})
shows a variety of features in the hot ICM, such as cold fronts and shocks, 
which trace the cluster formation 
history and provide invaluable information concerning their dynamical state 
(see Markevitch \& Vikhlinin \cite{mv07}). 
At the same time, the deep radio imaging achieved below 1 GHz by the 
Giant Metrewave Radio Telescope (GMRT), and more recently by LOFAR, is
shedding a new light on the properties of non-thermal components in galaxy
clusters.

Radio halos represent the most spectacular non-thermal effects of cluster 
mergers. Radio halos are diffuse synchrotron radio sources that cover the whole 
cluster volume, i.e. up and beyond Mpc size. These halos have steep spectra,
for which typical 
values are $\alpha\sim$ 1.2--1.3, for S$\propto\nu^{-\alpha}$, where S is the
flux density and $\alpha$ is the spectral index of the synchrotron radio
spectrum, and extremely 
low surface brightness of a fraction of $\mu$Jy arcsec$^{-2}$ (see the Feretti et 
al. \cite{feretti12} for a recent observational review).
Radio halos are not ubiquitous in galaxy clusters; only 30-40\% of massive 
clusters in the Universe (M$\ge10^{15}$ M$_{\odot}$) host a radio halo. 
In those cases, the 1.4 GHz radio power of the Mpc-scale halo 
correlates with both the cluster X-ray luminosity and mass
(e.g. Liang et al. \cite{liang00}, 
Brunetti et al. 
\cite{brunetti07}, Basu \cite{basu12}, Cassano et al. \cite{cassano13},
Cuciti et al. 
\cite{cuciti15}). For those clusters without a radio halo, upper limits to
the radio power derived from the GMRT radio halo survey are orders of magnitude
below the correlation (Venturi et al. \cite{venturi07} and \cite{venturi08}, 
Cassano et al.  \cite{cassano13}, Kale et al. \cite{kale13} and \cite{kale15a}).

The origin of giant radio halos is still a debated issue, however 
the connection between radio halos and cluster mergers is quantitatively 
supported by the distribution of clusters with and without radio halos as 
a function of a number of indicators of the X-ray substructure: giant 
radio halos are always found in unrelaxed clusters (Buote \cite{buote01}, 
Cassano et al. \cite{cassano10}).
On the other hand, a strong correlation is found between the cool-core 
strength in relaxed clusters and 
the presence of radio mini-halos; that is, diffuse cluster-scale sources that
are smaller in size than giant radio halos (of the order of up to few hundreds
of kpc), which always encompass the radio emission from the cluster dominant
galaxy, but whose origin is not closely related to the current cycle
of activity (e.g. Bravi et al. \cite{bravi16}, 
Giacintucci et al. \cite{giacintucci14}, Kale et al. \cite{kale15b}, 
Mittal et al. \cite{mittal09}).

A possible explanation for the origin of giant radio halos is the 
re-acceleration 
of in situ relativistic particles by turbulence injected into the ICM during 
cluster merger events (the so-called re-acceleration model; see Brunetti et al. 
\cite{brunetti01}).
Secondary (hadronic) models for the origin of the relativistic particles 
(e.g. Dennison \cite{dennison80}, Blasi \& Colafrancesco \cite{bc99}) are
currently less favoured because of the lack of detection of predicted 
$\gamma$-ray emission from galaxy clusters by the {\it Fermi} satellite,
and because of the discovery of radio halos with ultra-steep spectra ($\alpha \ge$ 
1.5), whose relativistic energy would exceed the thermal energy under the 
secondary model assumptions (see the prototype case of A\,521; Brunetti et al.
\cite{brunetti08}). 
Mixed hadronic and re-acceleration models have been proposed 
and some level of radio emission is expected in ``radio off-state clusters''
(e.g. Brown et al. \cite{brown11}, Brunetti \& Lazarian \cite{bl11}, 
Zandanel et al. \cite{zandanel14}).

Despite the statistical connection between radio halos and mergers, a few
radio halos have been found in less disturbed systems. An example of these
outliers is the giant radio halo in the strong cool-core cluster 
CL\,1821+643 (Bonafede et al. \cite{bonafede14}). A study of the X-ray 
morphological parameters of this cluster shows that it shares the same properties of 
galaxy clusters hosting a radio halo, suggesting that it may be a case
of a cluster merger in which the cool core has been preserved (Kale \& Parekh
\cite{kp16}).  More recently, a giant radio halo has been reported in
A\,2261 and A\,2390, neither of which is a major merger (Sommer et al.
\cite{sommer17}).
Current models predict that radio halos can also be generated in less 
disturbed systems, although with a probability that is significantly lower
than in the case of massive major mergers (Cassano et al. \cite{cassano06},
and Brunetti \& Jones \cite{bj14} for a review). Such outliers are thus
very important, as they may provide important constraints on
the origin of radio halos, and probe a piece of the theoretical framework
that is still poorly explored.
\\
\\
The galaxy cluster A\,2142 (z=0.0909) is another challenge to our
understanding of the formation of giant radio halos.

A\,2142 is massive (M$\sim 8.8\times10^{14}$ M$_{\odot}$, Cuciti et al.
\cite{cuciti15}) and is located at the centre of a supercluster
(Einasto et al. \cite{einasto15}, Gramann et al. \cite{gramann15}) with
ongoing accretion groups.
It is the first object where cold fronts were discovered by $Chandra$
(Markevitch et al. \cite{markevitch00}). Recently, another cold front
at the unprecedented distance of $\sim$ 1 Mpc from the  cluster centre
was detected with {\it XMM--Newton} and studied by Rossetti et 
al. (\cite{rossetti13}), who proposed that large-scale sloshing was the
responsible mechanism for its origin; this either resulted from the
long-term evolution of central sloshing common in
many relaxed clusters or from a merger of intermediate strength.
\\ 
Diffuse radio emission on the scale of few hundred kpc around the two
brightest cluster galaxies (BCGs) was detected by Giovannini \& Feretti 
(\cite{gf00}). A more recent study performed with the Green Bank Telescope 
(GBT) at 1410 MHz shows that the size of this diffuse radio emission is 
$\sim$2 Mpc, extending even beyond the most distant cold front. 
The major axis of this giant radio halo is aligned in the same south-east 
direction where the large-scale cold front is located 
(Farnsworth et al. \cite{farnsworth13}). The GBT image 
shows that its surface brightness is very low,  
i.e. $\sim 0.2~\mu$Jy arcsec$^{-2}$, but the poor angular resolution does not 
allow a detailed comparison with the X--ray images. To overcome the 
resolution limitations of the GBT image and to study the spectral properties
of this exceptional radio halo for comparison with the X--ray emission and
optical information, we undertook a study with the Giant Metrewave
Radio Telescope at 608 MHz, 322 MHz and 234 MHz, and with the Karl Jansky
Very Large Array (VLA) in the 1--2 GHz band. We complemented our analysis
with LOw Frequency ARray (LOFAR) data at 118 MHz and with archival VLA
observations at 1.4 GHz.

In this paper we present the results of our work. The paper is organized
as follows: in Sect. 2 we describe the observations and data analysis;
the images are presented in Sect. 3; the spectral study is presented in
Sect. 4; the results are discussed in Sect. 5; and the summary and future
prospects 
are given in Sect. 6. In the Appendix we complement the information and 
report on the radio emission associated with the cluster galaxies. 
Throughout the paper we use the convention 
S$\propto\nu^{-\alpha}$. We assume a standard $\Lambda$CDM cosmology with 
H$_o$=70 km~s$^{-1}$Mpc$^{-1}$, $\Omega_{\rm M}$=0.3, $\Omega_{\rm v}$=0.7.
At the cluster redshift (z=0.0909) this corresponds to a scale of 
1.705 kpc/$^{\prime\prime}$ and to a luminosity distance D$_{\rm L}$=418.6 Mpc.

\section{Observations and data reduction}\label{sec:obs}

\subsection{Observations with the GMRT}

We observed A\,2142 with the Giant Metrewave Radio Telescope (GMRT; 
Pune, India) in March 2013 at 608, 322, and 234 MHz. The logs of the 
observations are given in Table 1.


\begin{table*}[h!]
\caption[]{Observations}
\begin{center}
\begin{tabular}{clcccccc}
\hline\noalign{\smallskip}
Array & Project ID  &  Obs. Date & $\nu$   & $\Delta\nu$ & Time & FWHM & rms$^{(a)}$  \\ 
   & &       &    MHz &     MHz &  hour & $^{\prime\prime}\times^{\prime\prime}, ^{\circ}$& mJy~b$^{-1}$               \\
\hline\noalign{\smallskip}
GMRT & 23\_017  & 23-03-13 &  234  &  16  &  10 & 12.7$\times$11.0, 68& 0.18 \\
     & 23\_017  & 28-03-13 &  322  &  32  &  10 &  9.8$\times$7.5, 58 & 0.13 \\
     & 23\_017  & 22-03-13 &  608  &  32  &  10 &  5.4$\times$4.5, 59 & 0.03 \\
VLA$^{(b)}$ & VLA11B-156 & 09-10-11 & 1500 & 1000 & 1.5 & $^{(c)}$ & $^{(c)}$   \\
\hline\noalign{\smallskip}
\end{tabular}
\end{center}
Note: $^{(a)}$ Value measured far from the field centre; $^{(b)}$ the VLA 
observations consist of three pointings with the same set-up and exposure
time (see Sect. 2.2). Here we give the total time over the three pointings;
$^{(c)}$ see Sect. 2.2. and Table 2.
\label{tab:clusters}
\end{table*}


The data were collected in spectral-line mode at all frequencies, i.e. 256 
channels at 322 and 608 MHz, and 128 channels at 234 MHz, 
with a spectral resolution of 125 kHz/channel at 322 MHz and 608 MHz, and 
65 kHz/channel at 234 MHz. 
The raw data were first processed with the software {\it flagcal} 
(Prasad \& Chengalur \cite{prasad12}, Chengalur \cite{chengalur13}) to 
remove RFI and apply bandpass calibration, then further editing, 
self-calibration, and imaging were performed using the NRAO Astronomical 
Image Processing System (AIPS) package.
The sources 3C\,286 and 3C\,48 were used as primary (amplitude) calibrators. 
In order to find a compromise between the size of the dataset and the need to 
minimize bandwidth smearing effects within the primary beam, after bandpass 
calibration the central channels in each individual dataset were averaged to 
30, 39, and 26 channels of $\sim$1 MHz each at 608 MHz, 322 MHz, and 234 MHz,
respectively. 

\begin{figure*}[htbp]
\centering
\includegraphics[angle=0, scale=0.85]{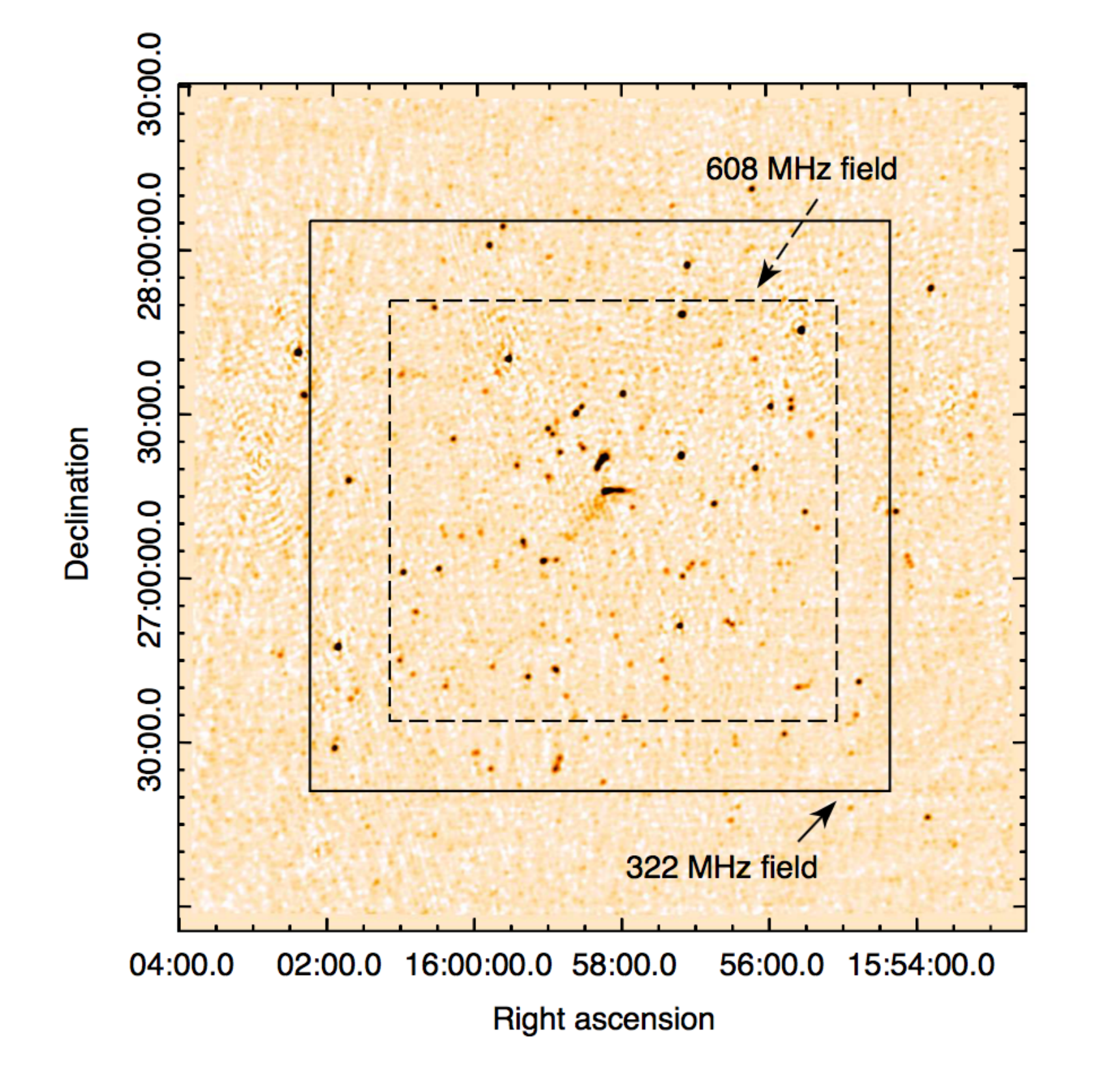}
\caption{Field of A\,2142 at GMRT 234 MHz . The low resolution image is restored with 
a beam of $45.2^{\prime \prime}\times 40.8^{\prime \prime}$, p.a. --37.9$^{\circ}$.
The continuous and dashed black boxes show the size of the 322 MHz and
608 MHz fields, respectively.}
\label{fig:fig1}
\end{figure*}

At each frequency we performed multi-facet imaging, covering an area
of $\sim2.5^{\circ}\times2.5^{\circ}$ at 234 MHz, $\sim1.8^{\circ}\times1.8^{\circ}$
at 322 MHz and $\sim1.4^{\circ}\times1.5^{\circ}$ at 608 MHz, respectively.
The field of A\,2142 is extremely crowded, as is clear from Fig. \ref{fig:fig1}, 
and includes many strong sources, which had to be properly self-calibrated 
and cleaned to reach the targeted rms (see Table 1).
\\
At each frequency we produced final images, over a wide range of resolutions
and with different tapering and weighting schemes, to account for the
complexity of the radio emission in the field. In particular, full 
resolution images were used to subtract the strongest radio sources at 
distances larger than $\sim 0.8^{\circ} - 1.5^{\circ}$ (depending on the 
frequency) from the field centre, and then tapering and robust weighting were 
used to image the diffuse radio sources and the radio halo.
Images with multiple resolutions were produced with resolutions
ranging from 
$5.5^{\prime\prime}\times4.5^{\prime\prime}$ to $39.1^{\prime\prime}\times36.6^{\prime\prime}$
at 608 MHz,
$9.8^{\prime\prime}\times7.5^{\prime\prime}$ to $53.4^{\prime\prime}\times43.7^{\prime\prime}$
at 322 MHz, and 
$12.7^{\prime\prime}\times11.0^{\prime\prime}$ to 
$45.19^{\prime\prime}\times44.80^{\prime\prime}$ at 234 MHz.

Finally, to image the radio halo we first produced high resolution images using 
the u--v spacings  $>2{\rm k}\lambda$. Then we used these to subtract the clean
components of the discrete sources from all of the u--v data, and imaged the 
residual emission with a taper and natural weighting (ROBUST=+2 in AIPS) 
to a resolution of $\sim50^{\prime\prime}-60^{\prime\prime}$ 
at all frequencies. All images were primary beam 
corrected using the task PBCOR in AIPS.
The shortest baselines in our datasets are $\sim$0.2 k$\lambda$,
  $\sim$0.1 k$\lambda,$ and $\sim$ 0.07 k$\lambda,$ respectively at 608 MHz,
  322 MHz, and 234 MHz. The largest detectable features are hence
17$^{\prime}$, 32$^{\prime}$ and 44$^{\prime}$, respectively.

The final rms values for the full resolution images are given in Table 1.
At lower resolution we obtained rms
$\sim$ 0.05 mJy~b$^{-1}$ at 608 MHz
($13.18^{\prime\prime}\times11.06^{\prime\prime}$), 
$\sim$ 0.25 mJy~b$^{-1}$ at 322 MHz
($14.80^{\prime\prime}\times13.06^{\prime\prime}$), 
and $\sim$ 0.3 mJy~b$^{-1}$ at 234 MHz
($19.46^{\prime\prime}\times17.92^{\prime\prime}$). 
We estimate that residual calibration errors at each frequency are 
within 4--5\% at 608 MHz and of the order of 10\% at 234 MHz and 322 MHz.
We point out that the difference in the flux density between the Baars et al. 
(\cite{baars77}) scale adopted here and the Scaife \& Heald (\cite{sh12}) 
scale, suggested for observations at $\nu\le 500$ MHz, is of the
order of few percent, hence within our final estimated errors.
Fig. \ref{fig:fig1} shows the full field of view at 234 MHz and highlights 
the size of the fields imaged at 322 MHz and 608 MHz.

%
%

\subsection{Jansky VLA Observations}

A2142 was observed with the Karl G. Jansky Very Large Array (VLA) in D and C
configurations at 1-2 GHz as part of NRAO observing programme VLA11B-156. 
Three pointings were acquired to recover the full extent of the radio halo
detected with the GBT 
(Farnsworth et al. \cite{farnsworth13})  with 28 
minutes of integration time per pointing. Observations were made in spectral 
line mode with 16 spectral windows, each 64 MHz wide, spread across the
full 1-2 GHz band. For technical reasons due to the recent VLA upgrade,
only two seconds of every five second integration on source were recorded, 
which resulted in higher thermal noise and less complete u-v coverage. 
Standard data flagging and reduction techniques were performed with CASA, 
using the VLA calibrator sources J\,1331+3030 (3C\,286) and J\,1609+2641 for 
flux and phase calibration, respectively.
After editing for RFI, roughly 45\% of the total bandwidth remained, yielding
$\sim$450 MHz over seven clean spectral windows. We created two images, one
using 250 MHz bandwidth around 1.38 GHz and one using 200 MHz around 1.78 GHz.
The standard phase and amplitude calibration were successful enough that
self-calibration did not produce a significant improvement, so it was not used
in the image presented here. Residual amplitude calibration errors are
estimated to be within 3\%.


\begin{table}[h!]
\caption[]{Parameters of the full resolution VLA images}
\begin{center}
\begin{tabular}{cccr}
\hline\noalign{\smallskip}
$\nu$   & $\Delta\nu$ & FWHM & rms  \\ 
  MHz   &     MHz     & $^{\prime\prime}\times^{\prime\prime}$ & mJy~b$^{-1}$ \\
\hline\noalign{\smallskip}
1380  & 250 & $40^{\prime\prime}\times37^{\prime\prime}$ & 150 \\
1780  & 200 & $32^{\prime\prime}\times29^{\prime\prime}$ &  70 \\
\hline\noalign{\smallskip}
\end{tabular}
\end{center}
\label{tab:vladata}
\end{table}


We used  the multi-frequency multi-scale clean task in CASA
(Rau \& Cornwell \cite{rc11}) both to deconvolve and create a mosaic from
the three pointings for each of the 1.38 and 1.78~GHz maps.
Correction for primary beam attenuation was performed using the CASA task 
{\it impbcor}.
\\
To isolate the diffuse cluster emission we subtracted the
contribution from radio galaxies, as follows.
We used the C configuration 1.6~GHz maps (resolution of 11$^{\prime\prime}$,
rms sensitivity of 90 $\mu$Jy~b$^{-1}$) to create masks of the location
and extents of radio galaxies; these data were not calibrated 
accurately enough to directly subtract from the D configuration data.
With these masks, we then performed an interactive single-scale clean of
the 1.38~GHz and 1.78~GHz full resolution $\sim$40$\arcsec$ D configuration
images until the radio galaxies were no longer visible.   
Model u--v datasets representing the radio galaxy emission were created
from these clean components and subtracted from the original D configuration
u--v data. The residual u-v datasets were then imaged with multi-scale clean
and corrected for the primary beam, convolving the final maps to
60$^{\prime\prime}$ with an rms of $\sim$180 $\mu$Jy~b$^{-1}$
($\sim140~\mu$Jy~b$^{-1}$) at field centre for 1.38 (1.78)~GHz, to increase
the signal-to-noise ratio of the diffuse emission.
The largest detectable angular size of these observations is 
\simlt$1000^{\prime\prime}$ and \simlt$820^{\prime\prime}$, at 1.38 GHz and 1.78
GHz, respectively.

\section{Radio emission from cluster galaxies}

The inner portion of the field of view of the GMRT observations, i.e. 
A\,2142 itself, is shown in Figs. \ref{fig:fig2} and \ref{fig:fig3},
where the contours of the 234 MHz and 608 MHz radio emission are 
overlaid on the red plate of the Digitized Sky Survey DSS--2  
and the {\it XMM--Newton} image, respectively. 

The central region of A\,2142 is dominated by the presence of two 
extended FRI radio galaxies (Fanaroff \& Riley \cite{fr74}) with head-tail 
morphology (Sect. 3.1) and by diffuse emission coincident with the 
brightest part of the X-ray emission from the intracluster medium, which we
classify as a giant radio halo (Sect. 4). 
Beyond these striking features, many radio sources in the field  
are associated with cluster galaxies.

The online table in the Appendix reports the full sample of radio sources 
with optical counterpart at the redshift of A\,2142 in the 608 MHz
field of view (Fig. \ref{fig:fig1}).
The list is based on the full resolution 608 MHz image with a detection 
limit S$_{\rm 608~MHz}$=0.25 mJy (i.e. 5$\sigma$) prior to the primary beam 
correction. For this reason, the radio source catalogue is not complete in 
radio power, whose detection limit increases away from the cluster centre. 
Flux density values at 234 MHz have also been reported in the table.
The radio galaxies presented in Sect. 3.1 are shown in Fig. A.1 
(Appendix), where the full resolution 234 MHz and 608 MHz 
contours are overlaid on a CFHT (Canadian French Hawaii Telescope) MegaCam
g-band image.

\subsection{Radio galaxies at the cluster centre}

The most striking radio galaxies at the centre of A\,2142 are two long 
tailed sources labelled T1 and T2 in Fig. \ref{fig:fig2}. A zoom on each 
of them is shown in the Appendix in the upper left and upper right
  panels of  Fig. A.1.

The source T1 is the radio galaxy B2~1556+27 (Colla et al. \cite{colla72}, 
Owen et al. \cite{owen93}), associated with a m$_{\rm g}$=17.5
cluster galaxy (z=0.0955). A compact counterpart is visible in the
{\it XMM-Newton}  image.
The head of this radio galaxy is coincident with the north-western 
cold front in the cluster.
The length of the tail is $\sim$ 610 kpc. Fig. A.1 clearly shows that
the long  and straight tail has small amplitude wiggles.
The imaging 
process at all frequencies reveals that it is embedded in the 
diffuse emission of the radio halo. Its radio power 
is logP$_{\rm 608~MHz}$=24.80 W~Hz$^{-1}$, which is high for this class
of objects.
We measure $\alpha_{\rm 234~MHz}^{\rm 608~MHz}=0.73\pm0.15$,
which steepens to $\alpha_{\rm 608~MHz}^{\rm 1400~MHz}=1.04\pm0.11$.
\footnote{The VLA C+D configuration 1.4 GHz observations are a
  re-analysis of project AG344. We produced images with angular resolution 
$24.1^{\prime\prime}\times21.9^{\prime\prime}$ and 
$\sim38.6^{\prime\prime}\times34.9^{\prime\prime}$ 
with rms of the order of $\sim 15\mu$J~y~b$^{-1}$ and $\sim 20~\mu$~Jy~b$^{-1}$, respectively. Individual source subtraction was performed following
the same procedure described in Sect. 2.1 to obtain images of the 
diffuse emission.}

The source T2  is associated with the galaxy 2MASX J\,15582091+2720010
(m$_{\rm g}$=16.6, z=0.0873 from the NASA/IPAC Extragalaxtic Database; NED),
located north of the cluster centre and outside
the brightest part of the X--ray emission, as is clear from Fig. \ref{fig:fig3}. 
Its length is $\sim370$ kpc and its radio power is logP$_{\rm 608~MHz}$=24.51
W~Hz$^{-1}$.
From Table A.1 it is clear that T2 has a very steep integrated spectrum. If we 
complement our GMRT flux density measurements with the archival 1.4 GHz data, 
we obtain $\alpha_{\rm 234~MHz}^{\rm 608~MHz}=1.21\pm0.14$,
which steepens to $\alpha_{\rm 608~MHz}^{\rm 1400~MHz}=1.92\pm0.11$.
 
Two wide-angle tail (WAT) sources are also present, and labelled W1 and W2 in
Fig. \ref{fig:fig2}. A zoom on each of them is shown in the Appendix in
the left and right central panels of Fig. A.1, respectively.
Interestingly, neither of these is located at 
the cluster redshift.
The source W1 has a very faint optical counterpart 
(m$_{\rm g}$=23 from NED) with 
z$_{\rm phot}=0.574$; its flux density is S$_{\rm 608~MHz}$=37.99 mJy. 
The radio peak of W2 coincides with an X-ray source and has a very faint 
optical counterpart. Despite the presence of three very nearby cluster 
galaxies, an association with any of these seems unlikely.
Considering that wide-angle tails are tracers of galaxy clusters (e.g.
Giacintucci \& Venturi \cite{gv09}; Mao et al. \cite{mao09}) and that both
W1 and W2 are associated with very faint objects, at least another 
cluster along the line of sight and behind A\,2142 must be present.

The radio emission labelled G (bottom left panel of Fig. A.1)
was presented in Eckert et al. (\cite{eckert14}) at 608 MHz.
It is a remarkable blend of discrete radio sources 
associated with a group at z$\sim$0.094 located north-east of the
cluster centre. The southern tip of the radio emission 
is coincident with the bright tip of the long ($\sim$ 800 kpc) 
X-ray tail visible in Fig. \ref{fig:fig3}, which has been
interpreted as the signature of the infall of the group into A\,2142 
(Eckert et al. \cite{eckert14} and  \cite{eckert17}). We associate the
southern peak of this emission with the brightest ($m_{\rm g}$=16.1) galaxy
in the group (Table A.1), even though the overlay shown in Fig. A.1
suggests that it could be a blend of emission from more galaxies.

The three compact radio galaxies labelled C1, C2, and C3 in  
Fig. \ref{fig:fig2} form another interesting group.
They are aligned and associated with cluster galaxies of 
similar optical magnitude (in the range m$_{\rm g}$=17.4--17.7 from NED).
A fourth cluster radio galaxy, C4, is located just west of this triplet,
as seen in the bottom right panel of Fig. A.1 in the Appendix.

The high resolution images at 608 MHz reveal that some cluster galaxies
embedded in the giant halo have associated radio emission. The 
most luminous BCG (m$_{\rm g}$=16.2 and z=0.0904) hosts a faint
radio source (see bottom left panel of Fig. \ref{fig:fig1app}).
Moreover, a fainter cluster galaxy 
(m$_{\rm g}$=18.8, z=0.0806) just north-west of the most luminous BCG 
and two more cluster galaxies located along the X-ray elongation from the
BCGs to the C1--C2--C3 group show radio emission.

\subsection{Overall radio properties of the galaxies in A\,2142}

A total of 42 radio sources have an optical counterpart.
These numbers refer to the region covered by the 608 MHz observations.
With exceptions made for 
T1 (GMRT~J\,155814+271619) and 
T2 (GMRT~J\,155820+272000),
the two tailed radio galaxies at the cluster centre, 
all radio sources are either point--like or barely extended at the full 
resolution of the 608 MHz image with radio powers in the range 
21.85 $\le$ logP$_{\rm 608~MHz}$ \simlt 23 W~Hz$^{-1}$.
These values suggest that the radio galaxy population of A\,2142
may include both starburst galaxies and faint radio active
nuclei. For the latter, the radio powers are well below the FRI/FRII 
division and are typical of the faint FR0 radio galaxies, whose radio 
morphology lacks extended emission in the form of radio lobes 
(Baldi et al. \cite{baldi15} and references therein). 

The radio emitting galaxies in A\,2142 span a range of redshifts from
$\sim$ 0.079 to \simgt 0.12 (see Table A.1);
this is consistent with the presence of multiple groups, as highlighted in the 
spectroscopic analysis performed by Owers et al. (\cite{owers11}) and
in the structure analysis shown in Einasto et al. (\cite{einasto15}).
In Fig. A.2 in the Appendix we show the location of the cluster radio 
galaxies listed in Table A.1 with colour codes to highlight the different 
redshifts. 

The  wider field of view of the 234 MHz (Fig. \ref{fig:fig1}) 
includes many more radio sources associated with cluster galaxies.
Starting from the visual inspection on DSS--2 and after consultation
of the NED database we found 71 counterparts in the redishift 
range z$\sim 0.07-0.13$. These are mostly located south and east of 
the cluster centre, following the overall elongation of the cluster
subgroups (Owers et al. \cite{owers11}) and of the supercluster
(Einasto et al. \cite{einasto15}). 
A full characterization of the population of radio sources in A\,2142 is 
beyond the scope of this paper and will be presented in a future work.

%
\begin{figure*}[htbp]
\centering
\includegraphics[angle=0, scale=0.8]{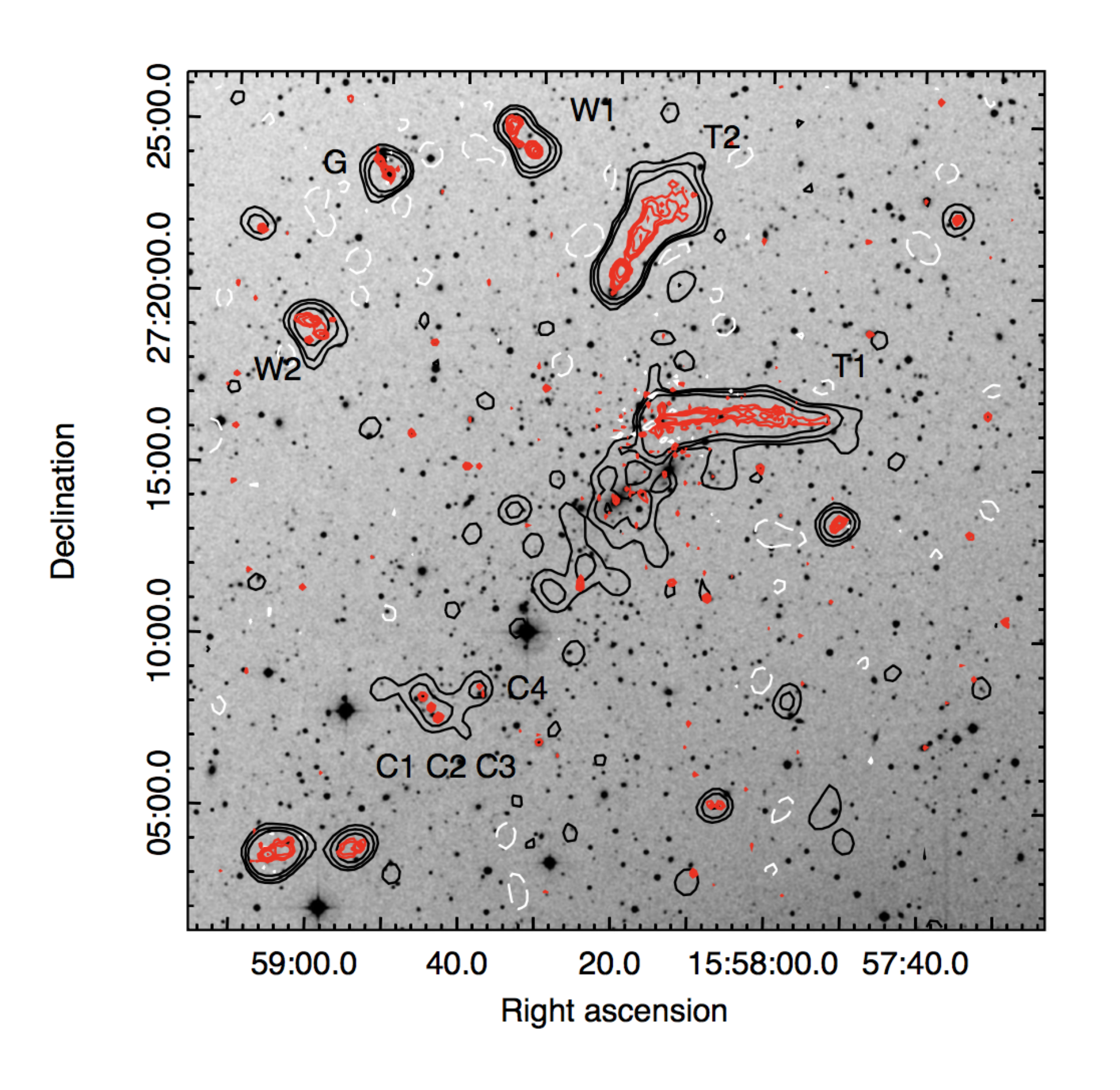}
\caption{Radio emission of A\,2142 overlaid on the red plate of DSS--2. 
Black contours show a low resolution GMRT 234 MHz image restored with 
a beam of $45.2^{\prime \prime}\times 40.8^{\prime \prime}$, p.a. --37.9$^{\circ}$ 
(same as Fig. 1); contour levels are $\pm$3,6,12 mJy/b; and the rms in the image
is $\sim$0.9 mJy~b$^{-1}$ far from the field centre (negative contours are
white dashed). 
Red contours show the 608 MHz full resolution image restored with a
beam of $5.2^{\prime \prime}\times 4.5^{\prime \prime}$, p.a. 52.6$^{\circ}$;
contour levels are $\pm$0.1,0.2,0.4,0.8,1.6,6.4,25.6,102.4 mJy~b$^{-1}$;
and the rms in the image is $\sim35\mu$~Jy~b$^{-1}$ far from the field centre.
Negative contours are shown in white.}
\label{fig:fig2}
\end{figure*}

%
\begin{figure*}[htbp]
\centering
\includegraphics[angle=0, scale=0.8]{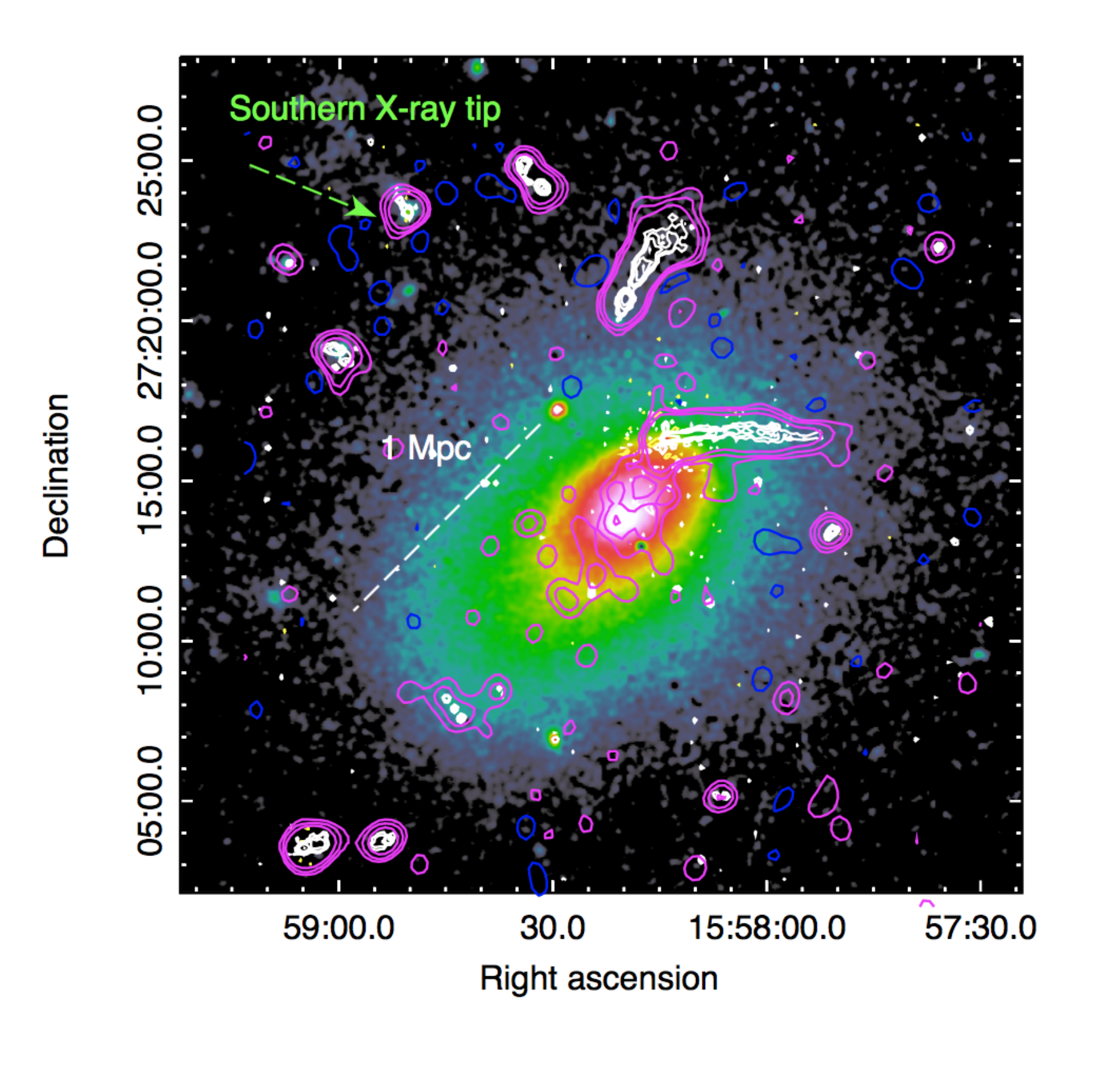}
\caption{Radio emission of A\,2142 overlaid on the X-ray image
from $XMM$-$Newton$ . Magenta contours show a 234 MHz low resolution 
image restored with a beam of $60^{\prime \prime}\times 60^{\prime \prime}$.
Contours are drawn at 1.5,3,6,12,24,48 mJy~b$^{-1}$. Negatives contours are
shown in blue (-3mJy~b$^{-1}$). The white contours show the
608 MHz image (same contours and resolution as in Fig. 2).}

\label{fig:fig3}
\end{figure*}

\section{Radio halo} 

\subsection{Morphology}

\begin{figure*}[htbp]
\includegraphics[angle=0, scale=0.35]{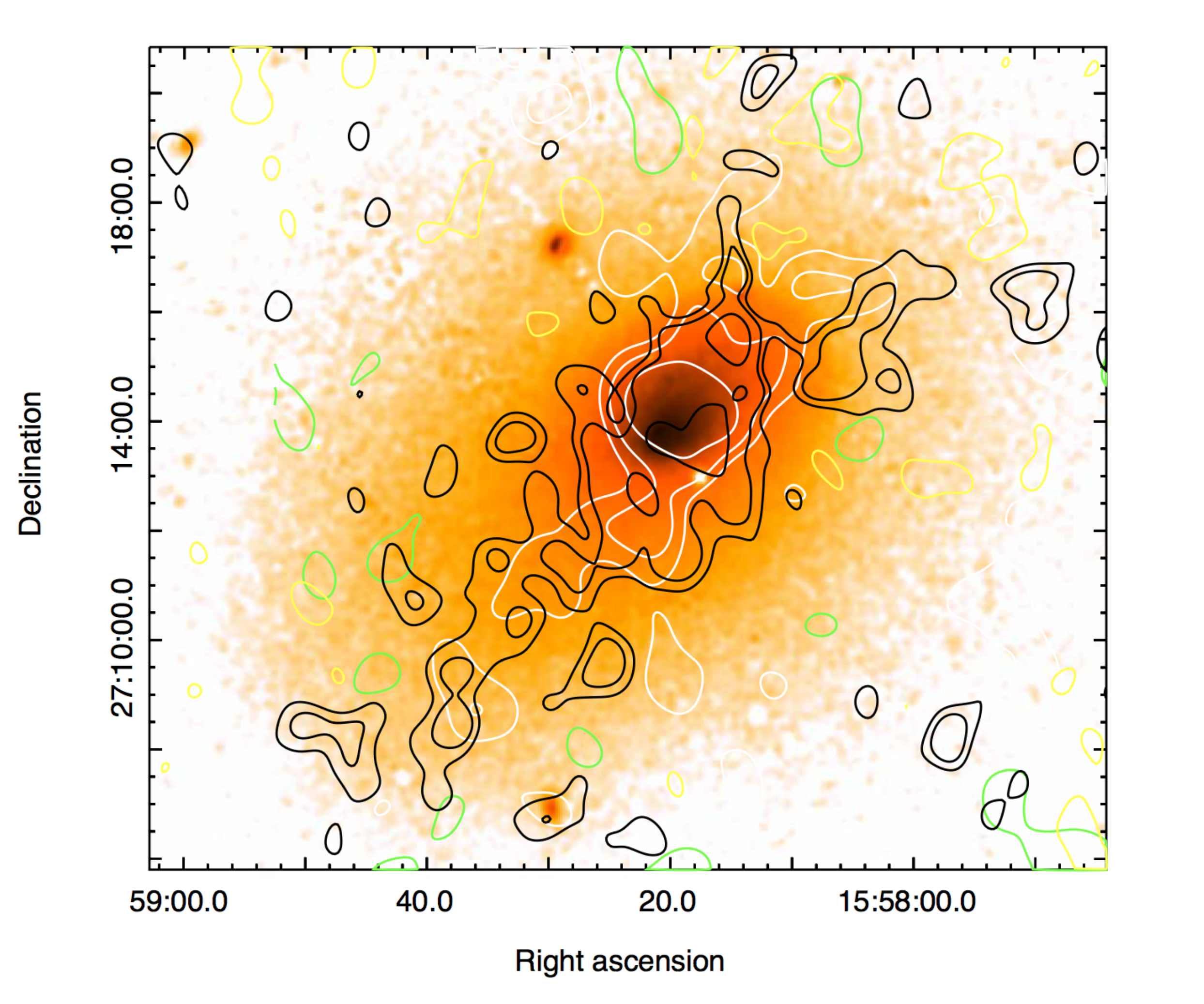}
\hspace{0.15truecm}
\includegraphics[angle=0, scale=0.42]{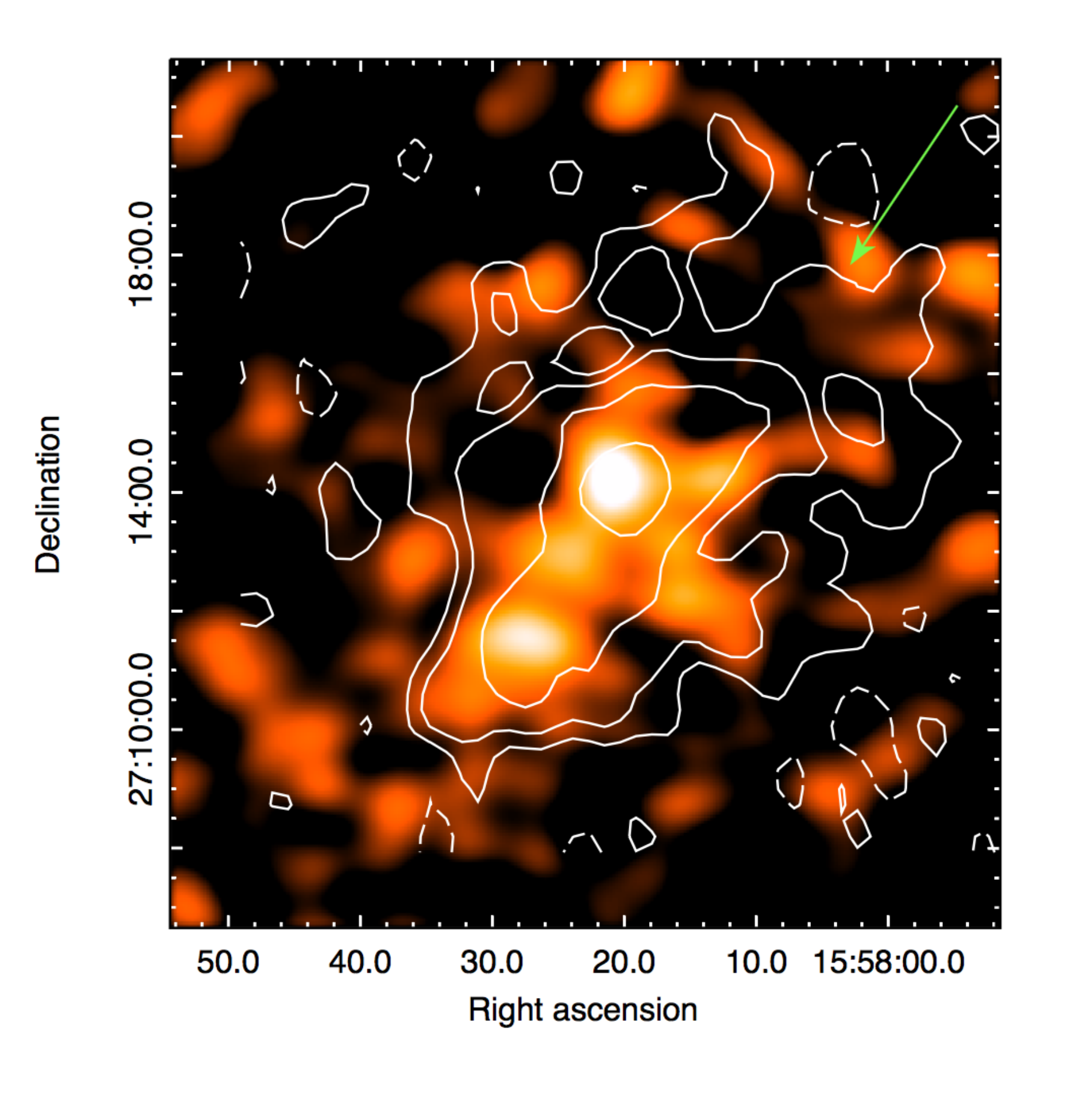}
\caption{{\it Left panel}: Radio halo in A\,2142 overlaid on  the 
{\it XMM--Newton} emission. The 234 MHz image is shown in black (negative 
contours in yellow). Contour levels are drawn at $\pm 2.5$, 4, 8 
mJy~b$^{-1}$, $\theta=44.86^{\prime\prime}\times40.96^{\prime\prime}$,
p.a. $-39^{\circ}$. The 608 MHz emission is shown in white (negative 
contours shown in green). Contour levels are drawn at $\pm$0.6, 1.2, 2.4 
mJy~b$^{-1}$, $\theta=50^{\prime\prime}\times50^{\prime\prime}$. Both
images were obtained after subtraction of the discrete sources
from the u--v data. 
{\it Right panel:} VLA contours at 1377 MHz 
($\theta=60^{\prime\prime}\times60^{\prime\prime}$, contours are drawn at
$\pm$0.2, 0.4, 0.8, 1.6 mJy~b$^{-1}$) overlaid on the 322 MHz GMRT image.
Negative contours are drawn as dashed lines. The green arrow highlights the north-western
extension (see Sect. 4.1).}   
\label{fig:fig4}
\end{figure*}

\begin{figure}[htbp]
\centering
\includegraphics[angle=0, scale=0.35]{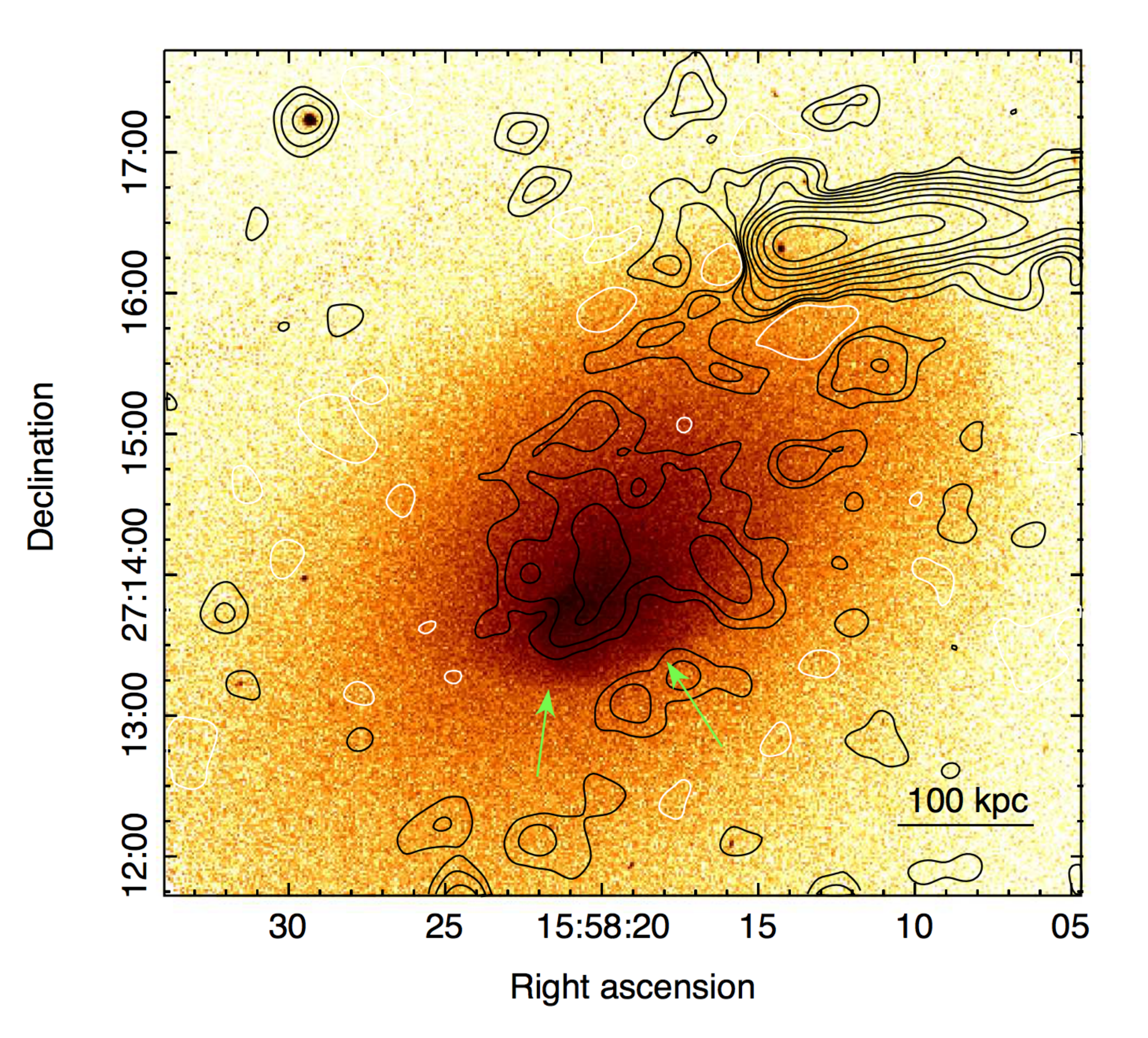}
\caption{Zoom on the central part of the radio halo. The GMRT 608 MHz image 
  (discrete radio sources not subtracted) at the resolution of
  $16.3^{\prime\prime}\times 14.2^{\prime\prime}$, p.a. $-84.6^{\circ}$, is overlaid
  on {\it Chandra}. Contours are drawn at $\pm$0.18, 0.36, 0.72, 1.44, 2.88
  mJy~b$^{-1}$; positive contours are shown in black, negative contours white). The green
  arrows show the location of the inner SE cold front.}
\label{fig:fig5}
\end{figure}

The diffuse extended emission in A\,2142 is well visible in Figs. 
\ref{fig:fig1}, \ref{fig:fig2}, and \ref{fig:fig3}, and this emission is best 
highlighted in the low resolution images shown in Fig. \ref{fig:fig4}
(obtained after subtraction of the individual radio galaxies at each
frequency; see Sects. 2.1 and 2.2), and in Fig. \ref{fig:fig5}.
Our images are suggestive of a multi-component cluster-scale emission,
which all together we refer to as  ``the radio halo''.
For our study we identify two regions, which are shown in the left
panel of Fig. \ref{fig:fig6} and are named H1 and H2.
The operational definition of these two regions is given in Sect. 4.2.

The region H1 is the brightest part of the halo, and this region is best
highlighted in Fig. \ref{fig:fig5}, which shows an intermediate resolution
608 MHz image overlaid on the $Chandra$ X-ray emission. The radio and X-ray peaks
are coincident. This region was  formerly classified as mini--halo
(Giovannini \& Feretti \cite{gf00}). 
Fig.  \ref{fig:fig5} clearly shows that it is confined by the innermost
cold front, whose position is indicated by the green arrows. Even though the 
north-western boundary is more difficult to define, owing to the presence of 
the radio emission from T1, 
H1 does not seem to extend beyond the north-western cold front.
This region has the same extent and boundaries at all frequencies.

The left panel of Fig. \ref{fig:fig4} shows the entire extended
radio emission in A\,2142 at 234 MHz and 
608 MHz\ overlaid on the X-ray emission detected by {\it XMM-Newton}, 
while the emission at 1.38 GHz is given as contours in the right panel, 
overlaid with the 322 MHz image.
At all frequencies, the radio halo is elongated in the same north-west 
to south-east direction of the X-ray emission from the intracluster gas, 
and covers the brightest X-ray ridge of emission out to the most distant 
cold front. Its largest angular size is $\sim 10^{\prime}$, 
i.e. $\sim$1 Mpc at the cluster redshift.
We define as H2 the ridge-like emission extending from H1 towards the
most distant old front in the south-east direction. Inspection of both panels
of Fig. 4 suggests that the surface brightness distribution of this component
is very different from H1. It is considerably more extended than H1 with a
lack of a central peak. The difference between H1 and H2 is confirmed by our
analysis of the radio brightness profile across the two regions shown in the
left panel of Fig. 6 (see Sect. 4.2 for details). The region H1 shows a regular
profile with a prominent peak, following closely the brightness distribution
of the hot gas. The profile of H2 is less regular, with a flat shoulder and
a steeper profile far from the core, and shows no clear correlation with
the X-ray profile. The north-western extension visible at 1.38 GHz
(right panel of Fig. \ref{fig:fig4}) is not detected at lower frequencies
and it is most likely due to incomplete subtraction of T1.

Finally, our VLA datasets detect further extended emission surrounding H1
and H2 (see the left panel of Fig. \ref{fig:fig6}), which we interpret as the 
remains of the 2 Mpc scale emission imaged with the GBT (Farnsworth et al.
\cite{farnsworth13}), whose full extent is unrecovered in all our images.
This clearly shows the limitations of interferometric
observations, whose lack of zero spacings is a severe limit in imaging
very low brightness extended and complex radio emission (as 
is the case of the radio halo in A\,2142) at least out to z$\sim$0.1.

In the next Sections we refer to H1 as the central emission, 
H2 as the ridge, and the larger scale emission for the more
diffuse emission detected with the VLA and GBT.

\subsection{Radio spectrum}

\begin{figure*}[htbp]
\centering
\includegraphics[angle=0, scale=0.32]{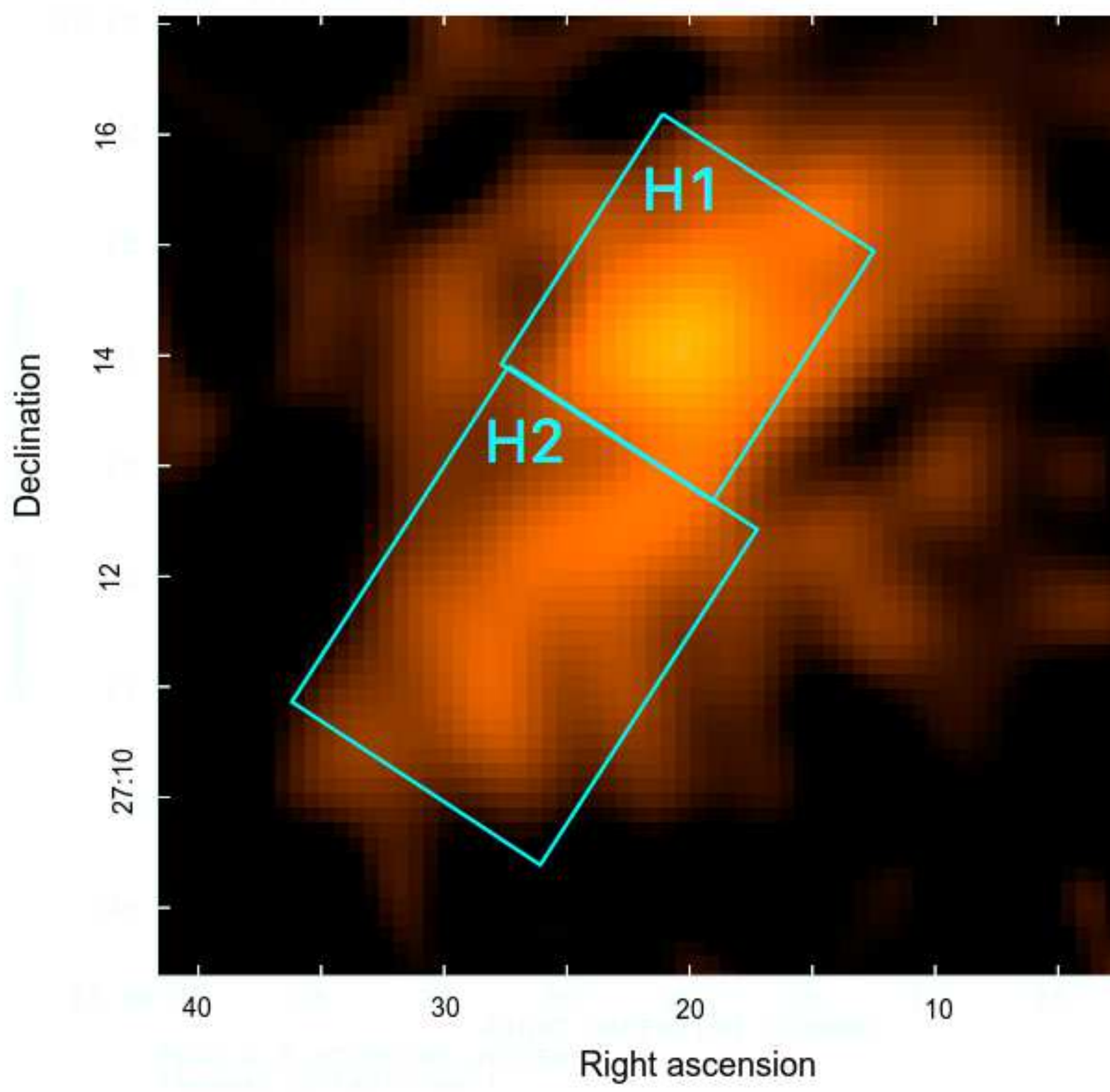}
\hspace{0.5truecm}
\includegraphics[angle=0, scale=0.25]{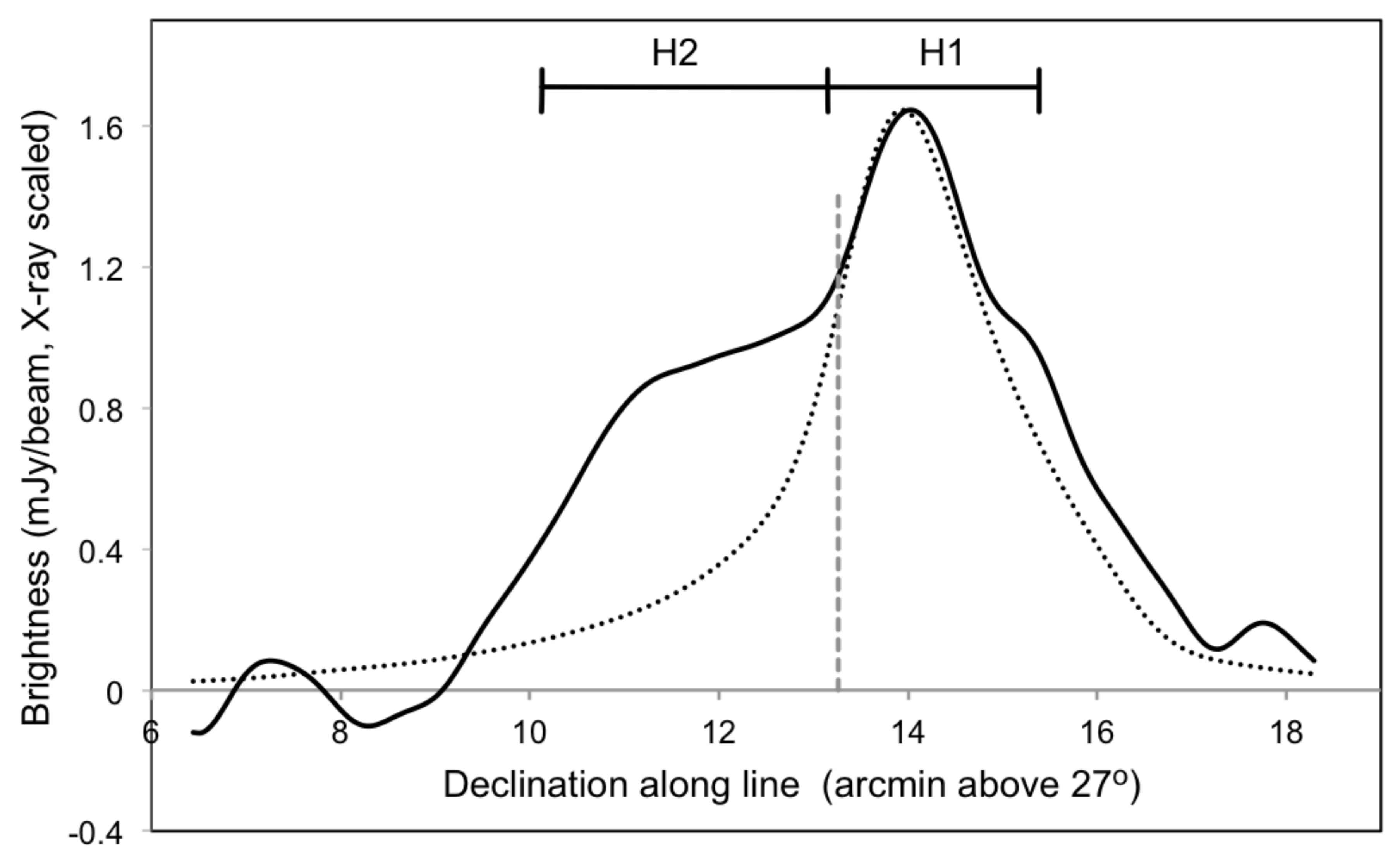}
\caption{{\it Left panel:} Two regions H1 and H2 used for the 
evaluation of the integrated spectrum (see Sect. 4.2) shown on the
1.38 GHz VLA image at the resolution of $60^{\prime\prime}$.  
{\it Right panel:} Slice of the surface brightness of the 1.38 GHz image
along the direction $-33^{\circ}$ (see Sect.  4.2), so that lengths
along the slice can be calculated as (Dec2- Dec1)/cos(33d).
The vertical grey dotted line shows the location of the south-east
innermost cold front, while the XMM X-ray brightness profile convolved
with the same resolution (60$^{\prime\prime}$) is shown as a black dotted
line.}
\label{fig:fig6}
\end{figure*}

The complexity of the radio emission in A\,2142 at all frequencies
does not allow reliable imaging of the spectral index distribution 
throughout the radio halo. To overcome this difficulty and obtain some 
information about possible changes of the spectral properties, we 
derived the integrated spectrum of the radio halo in H1 and H2 
(see the left panel of Fig. \ref{fig:fig6}, where the two
regions are overlaid on the VLA image).

To complement and further extend the frequency coverage of the GMRT and 
VLA observations we re-analysed 1.4 GHz archival data in the C and D 
configuration (see Sect. 3.1) and used a preliminary image obtained with 
LOFAR at 118 MHz.
The LOFAR \textsc{hba\_dual\_inner} 118-190\,MHz
data were recorded on April 19 2014 (project ID LC1\_017). A subset
  of the target dataset in the band 118-124\,MHz was calibrated and
  imaged using the standard direction independent calibration procedure
  (see e.g. Shimwell et al. \cite{shimwell17}). The resulting image,
  made from the visibilities from baselines shorter than 7\,k$\lambda$,
  has a sensitivity of 3~mJy~b$^{-1}$, an angular resolution of
  $\sim50\arcsec$ and the peak flux density measurements of compact
  sources are in agreement with the 150\,MHz TGSS ADR (Intema et al.
  \cite{intema17}) survey to within 30\%. A full direction dependent
  calibration of the dataset (see e.g. van Weeren et al. \cite{vanweeren16a},
  van Weeren et al. \cite{vanweeren16b}, Williams et al. \cite{williams16},
  Hardcastle et al. \cite{hardcastle16} and Shimwell et al. \cite{shimwell16})
  will be applied in the future to correct for ionospheric disturbances and
  improve the image fidelity, sensitivity and resolution, but this is beyond
  the scope of this paper.

Finally, we inspected the 74 MHz image in the VLSS--Redux (VLA Low-Frequency 
Sky Survey) to check for the presence of diffuse radio emission at this 
frequency and found no clear emission above the noise level.


To determine the spectral indices for H1 and H2, we carried
out the following steps: (a) we defined the H1 and H2 boundaries;
(b) we adopted a 
procedure that was insensitive to variations in the detailed structures at
different frequencies and that removed the
varying amounts of flux from the 2~Mpc components to measure the flux density of each component; and (c) we evaluated 
the u-v coverage to ensure reliable flux estimates.
We describe each of these below.

\begin{itemize}

\item{a)} H1 and H2 definition - The extents of H1 and H2 were initially
estimated by eye from greyscale images of the best images. To look at this
more quantitatively, we calculated the average flux at each position along
the major axis of the radio emission in a strip 160$^{\prime\prime}$ wide
along the minor axis.
The slice of the surface brightness, obtained using the VLA 1.38 GHz
image convolved to 60$^{\prime\prime}$, is shown in the right panel of Fig. 6.
Because H1 and
H2 overlap, the division between them is somewhat arbitrary. It was first
chosen as where H1 begins to significantly affect the shape of the H2
profile and then compared to the position of the cold front.
The H1, H2 division is at the southern base of the cold front (see Fig. 5
and right panel of Fig. 6).
The north-west boundary of H1 is approximately at its half-power level, as
is the south-east boundary of H2. Because of the variations in the detailed
structure at the different frequencies,
the widths of the H1 and H2 regions were chosen to be the FWHM of the
emission after smoothing by 160$^{\prime\prime}$ along the major axis. For H1,
we used the width of the bright core, rather than the lower brightness emission. The H2
profile is somewhat asymmetric, so the larger (SW) width was used.
These procedures led to the length and width of the boxes as
160$^{\prime\prime}\times136^{\prime\prime}$ (6 square arcmin) for H1 and
216$^{\prime\prime}\times160^{\prime\prime}$ (9.6 square arcmin) for H2,
oriented at $-33\deg$.

\item{b)} Flux density measurements - All images were first convolved to a
  resolution of 60$^{\prime\prime}$, except for the VLSSr, which has a beam size of
75$^{\prime\prime}$.
The total flux densities in the H1 and H2 boxes were then measured as follows. For
each component, we calculated the running sum of the flux within a box fixed
to the same length and width as H1 and H2, respectively. The box was slid
along the line perpendicular to the major axis over a length of
1120$^{\prime\prime}$ crossing the major axis (and H1 and H2, respectively)
near the middle. We then measured both the off-source background level in
each running sum, along with the peak flux above the background. These peak
fluxes are reported in Table 3. The background level includes any small
contributions from the much larger scale (20$^{\prime}$ component detected in
our single dish measurements), as well as any instrumental effects in the
interferometer images.
We then calculated the residual rms scatter in the off-source background of
each running sum and report these as the errors for the peak fluxes. Such
errors therefore reflect random noise and instrumental artifacts but do not reflect any
systematic effects such as those due to u-v coverage, which we discuss below.
It is important to note that the H1 and H2 fluxes are not the ``total flux''
from each component, since they each have broad wings beyond their box widths.
It is not clear whether these wings are extensions of H1 and H2 or part of the
larger scale emission. However, since the boxes were kept the same at all
frequencies, they provide a well-defined sample of the H1 and H2 emission.
The right panel of Fig. 6 shows the surface brightness profile along the ridge
oriented at $-33\deg$. The clear difference in surface brightness and
extent between H1 and H2 separately motivated our study of the total spectrum for
these two regions.

\item{c)} Evaluation of the u--v coverage - 
Despite the nominal largest detectable angular size of our observations at
each frequency (see Sect. 2.1 and 2.2, for completeness note that the
largest angular size detectable by LOFAR at 118 MHz is $\sim1^{\circ}$),
because of their very low surface brightness none of our interferometer data
sample well the largest scale (20$^{\prime}$) emission seen
with the single dish. We note that the total flux density detected with the
VLA at 1377 MHz is 23$\pm2$ mJy, which includes H1, H2, and the emission
beyond these two regions until they fade into the background, while the
1410 MHz flux density measured in the GBT image is $\sim 55$ mJy
(see Farnsworth et al. \cite{farnsworth13}).
However, the partial sampling of this very diffuse
component, which varies from frequency to frequency, can lead to elevated
background levels around H1 and H2. The determination of a background level
in the 1-D cuts, as discussed above, removes any such effects.
The total spectrum of H1 and  H2 between 118 MHz and 1.778 GHz is
shown in the left and right panel of Fig. 7, respectively.
We point out that H2, whose largest extent along its major axis 
(216$^{\prime\prime}$) is well within the nominal largest detectable structure, 
is not sufficiently well sampled in our interferometric images owing to to its
surface brightness, which is considerably lower than H1. 
In particular, the inner portion of the u--v coverage (within
1 k$\lambda$) is much better sampled at 118 MHz (LOFAR), 322 MHz (GMRT),
1.38 GH, and 1.78 GHz (VLA-D). For this reason we refer to those images
as to the well-sampled (we are using this or good coverage)
images.
By contrast, the u--v coverage of the 1.4 GHz archival VLA and the 608 MHz
and 234 MHz GMRT data is poorer on the shortest spacings and some of the
flux density from H2 is missing. This problem is well known and its effect
has been quantified in earlier works on giant radio halos (see Brunetti et al.
\cite{brunetti08} and Venturi et al. \cite{venturi08}).
The flux density measurements for H2 are plotted
separately in the right panel of Figure 7.
\end{itemize}


A weighted mean square fit provides 
$\alpha_{\rm 118~MHz}^{\rm 1.78~GHz}=1.33\pm0.08$ for H1 and
$\alpha_{\rm 118~MHz}^{\rm 1.78~GHz}=1.42\pm0.08$ for H2.
The difference is only marginal. However, it becomes more significant
if we take the different u--v coverage into account. A fit of the two sets
of measurements for H2 separately (filled green and filled blue dots
in the right panel of Fig. 7), provides 
$\alpha_{\rm 118~MHz}^{\rm 1.78~GHz}=1.55\pm0.08$.
The flux density upper limits at 74 MHz are consistent with the
spectral index both for H1 and H2. The future analysis of
the LOFAR data will allow us to obtain better constraints.


\begin{table}[h!]
\caption[]{Radio halo flux density}
\begin{tabular}{lrr}
\hline\noalign{\smallskip}
Region  &  Frequency & Flux density \\
        &    MHz     &  mJy         \\
\hline\noalign{\smallskip}
H1  & 1778 &   4.4$\pm$  0.5 \\
    & 1465 &   7.0$\pm$  0.5 \\
    & 1377 &   7.9$\pm$  0.4 \\
    &  608 &  16.5$\pm$  1.4 \\
    &  322 &  56.2$\pm$  4.1 \\
    &  234 &  69.3$\pm$  4.9 \\
    &  118 & 275.0$\pm$ 94.1 \\
    &   74 & $<$ 700   \\
\hline\noalign{\smallskip}
H2  & 1778 &   3.3$\pm$  0.5 \\
    & 1465 &   2.2$\pm$  0.4 \\
    & 1377 &   4.9$\pm$  0.4 \\
    &  608 &   6.8$\pm$  1.4 \\
    &  322 &  40.5$\pm$  4.1 \\
    &  234 &  34.5$\pm$  4.9 \\
    &  118 & 260.0$\pm$ 94.1 \\
    &   74 &  $<$ 700   \\
\hline\noalign{\smallskip}
\end{tabular}
\label{tab:halo}
\end{table}


%

%

%

%
\begin{figure*}[htbp]
\includegraphics[angle=0, scale=0.33]{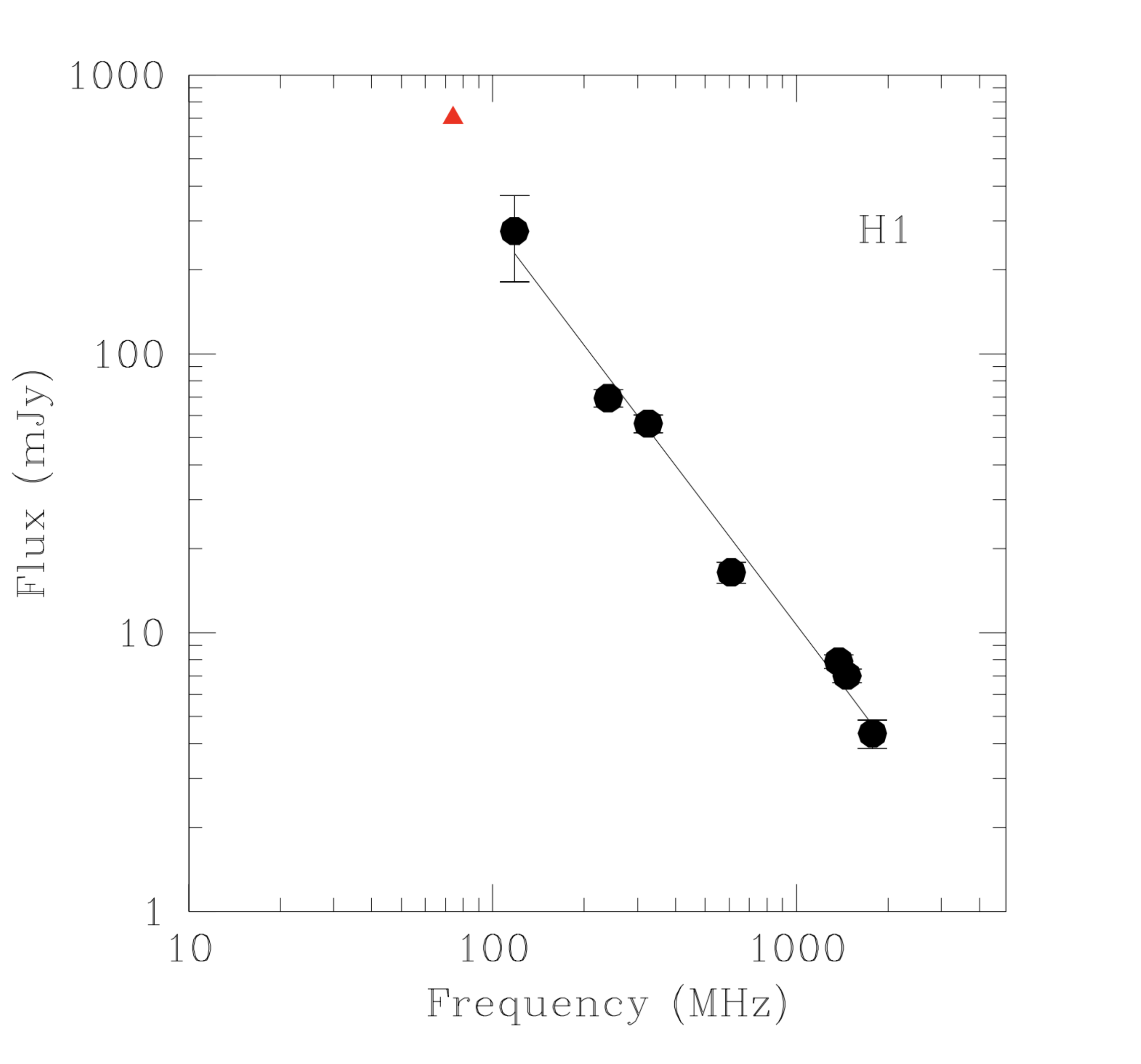}
\hspace{0.2truecm}
\includegraphics[angle=0, scale=0.33]{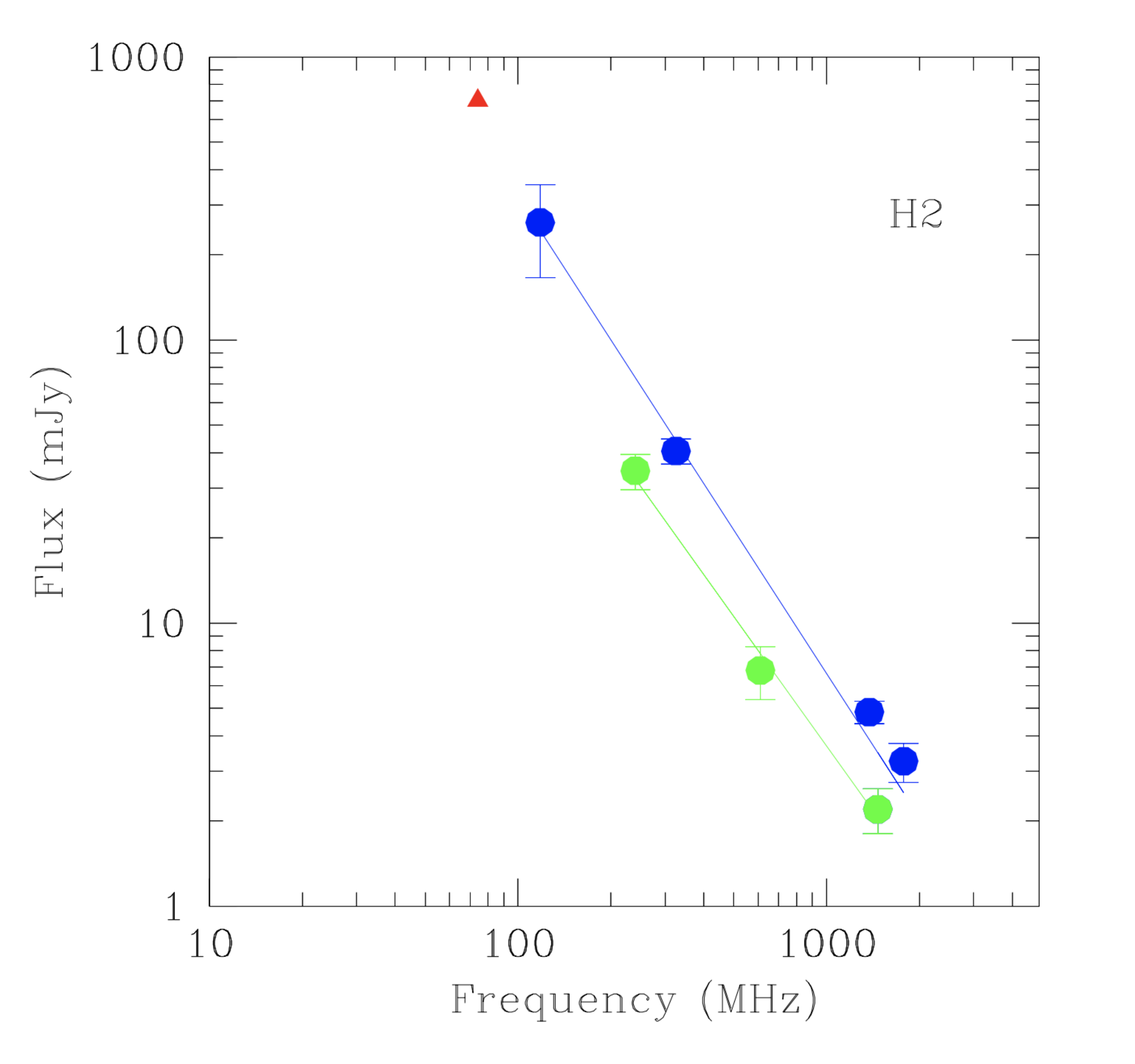}
\caption {Integrated spectrum of the radio halo. {\it Left panel:}
  Plot of H1. The upper limit at 74 MHz is shown as a red triangle.
  {\it Right panel:} Plot of H2. Filled blue circles highlight the
  measurements with good u--v coverage at short spacings; the remaining
  measurements
  are shown as filled green dots see Sect. 4.2). In both panels
  the weighted linear fits are shown with the same colour code and 
  the upper limit at 74 MHz is shown as a red triangle.}
\label{fig:fig7}
\end{figure*}

\section{Origin of the radio halo} 

The observations presented in this work confirm that in many respects A\,2142 is a case 
study of galaxy clusters. 
Our most important results is the finding that
the radio halo consists of two regions with different morphological
and spectral properties.

Our analysis, which spreads over a frequency range of more than one order
  of magnitude (118 MHz -- 1.78 GHz), suggests that the region of emission
  extending south-east of the cluster core (H2; see left panel of Fig. 6)
  has a lower surface brightness, is broader in size, and
  has a moderately steeper spectrum than the brighter, more compact
region in the core (H1).
In this Section we discuss the possible origin of such differences in the
framework of the cluster dynamics, as inferred from its broadband properties.
In Fig. 8 we show all the observational information that is relevant
to the discussion.

%
\begin{figure*}[htbp]
\centering
\includegraphics[angle=0, scale=0.7]{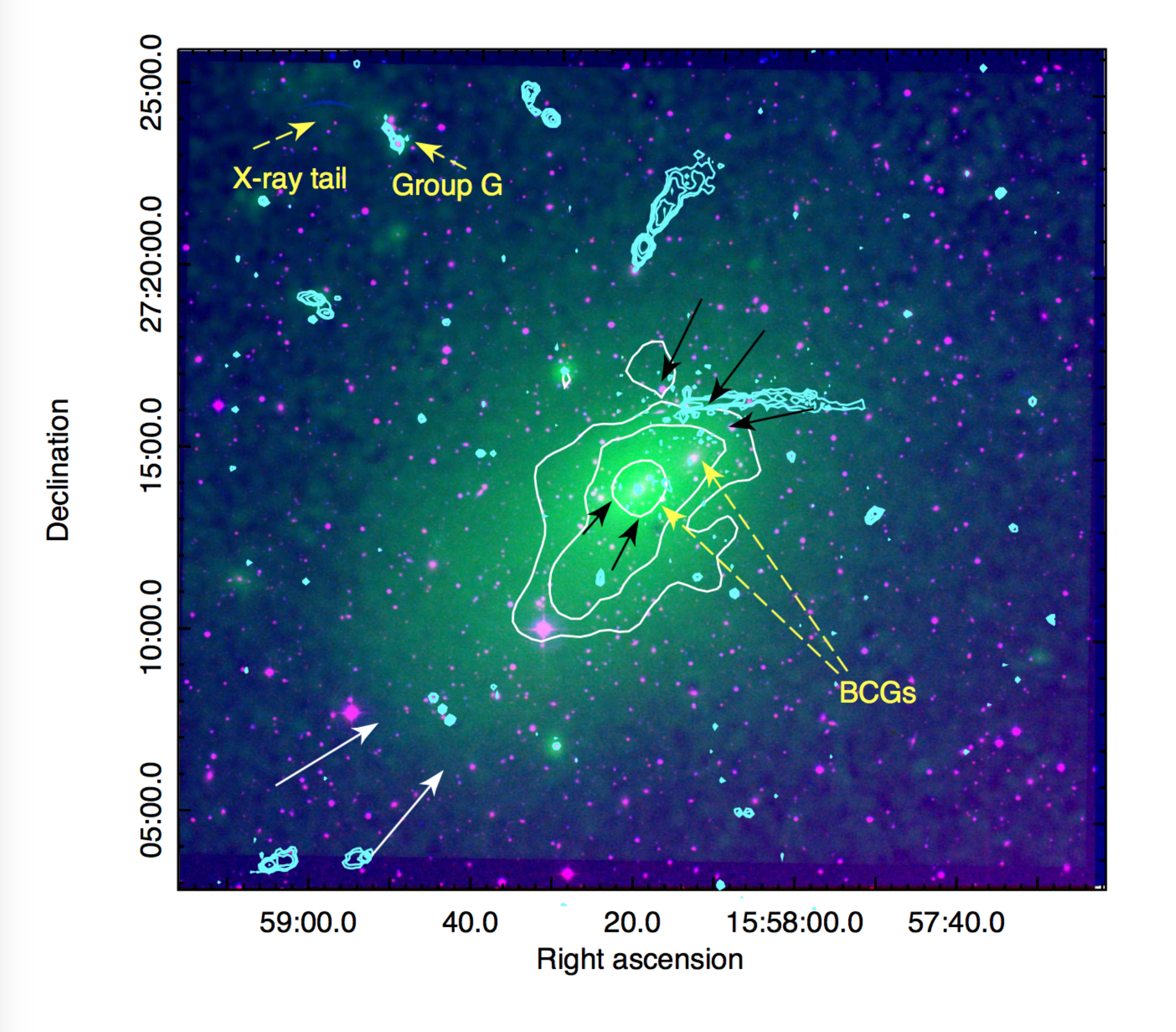}
\caption{Multi-band image of A\,2142. The optical information from the
  combined red and blue plates of DSS--2 is shown in purple, the X-ray
  emission from {\it XMM-Newton} is shown in green, radio contours at
  608 MHz are shown in cyan (same image and contour levels as in Figs. 2
  and 3), the emission from the radio halo at 1.38 GHz is shown as white
  contours (same image as in the right panel of Fig. 6, contour levels
  drawn at 0.4, 0.8, 1.6 mJy/b).
  The three cold fronts are indicated by the black
  and white arrows. The innermost cold front is coincident with
  the highest radio contour of the radio halo and the northernmost
  cold front is coincident with the lowest radio contours. For completeness,
  the location of the BCGs, of Group G and the X-ray tail discussed
  in Eckert et al. \cite{eckert14} are also shown.}
\label{fig:fig8}
\end{figure*}

\subsection{The central emission}

The radio emission observed in the core of A\,2142 (H1) 
is bounded by the two inner 
cold fronts that are detected on the 100-200 kpc scale, as clear
from Fig. \ref{fig:fig5} and \ref{fig:fig8}.
This may suggest that this emission traces the dissipation of the energy 
produced by the sloshing of the low-entropy gas oscillating between the two 
inner cold fronts. This scenario is similar to what has been suggested 
for the origin of mini--halos in cool-core clusters (e.g. Mazzotta 
\& Giacintucci \cite{mg08}, ZuHone et al. \cite{zuhone13}). 
Two BCGs are present in A2142 (see Figs. 8 and A.1) and the sloshing
in the core may be due to the perturbation induced by a gasless minor merger
with the group associated with the secondary BCG (Owers et al. \cite{owers11}).
These two BCGs may also play a role in the origin of the radio emission. 
At present only the most luminous BCG hosts a radio source (Sect. 3.1,
Table A.1 and Fig. A.1), but it is likely that over the last Gyr or so
both BCGs have released relativistic particles  and magnetic fields in 
the core region. During this timescale relativistic 
electrons could be advected by turbulent motions in the sloshing gas. 
The very weak correlation between the radio power of the BCG and that
of the mini-halo found for a sizeable sample of mini-halo clusters 
(Govoni et al. \cite{govoni09}, Giacintucci et al. \cite{giacintucci14}) is 
indeed suggestive of the
fact that the central AGN activity may not be powering the radio emission
directly, but it is the most likely source of seed electrons for 
re-acceleration in the ICM.

In order to cover a distance of the order of 100 kpc in 1 Gyr, the 
spatial diffusion coefficient due to turbulent transport should be of
the order of 
$D \sim \delta V_L L \sim 2 \times 10^{30}cm^2 s^{-1}$
(Brunetti \& Jones \cite{bj14} and references therein),
implying a velocity of the turbulent eddies of $\sim$100 km$/$s on a
scale of about 50 kpc. These requirements 
are indeed consistent with those measured in simulations of  gas 
sloshing and those necessary for turbulent re-acceleration to maintain 
(or to re-accelerate) radio emitting electrons in these regions (ZuHone et al. 
\cite{zuhone13}); in addition, values of the same magnitude range have been
derived for the Perseus cluster (Hitomi collaboration \cite{hitomi16}).

\subsection{The ridge}

The right panel of Fig. 6 clearly shows that the surface brightness and
extent of the radio emission in the ridge (H2) are very different from
those in the core region (H1). The clear separation between these
two components suggests that they might originate from different
mechanisms or that they trace different evolutionary stages of the
same phenomenon.
On the scale of the ridge the cluster appears unrelaxed with a clear
elongation in 
the south-eastern direction. The radio emission follows the spatial 
distribution of the X-ray emitting gas and extends up to the cold front  
located at 1 Mpc south-east of the core.
On these scales the cluster may have been perturbed by a number of
processes that induce gas sloshing and generate the external cold front.
The detection of a bridge of low-entropy gas between the central region
of the cluster and the most distant cold front is in support of this picture
(Rossetti et al. \cite{rossetti13}).
It is likely that the gas and magnetic fields
in this region have been moderately perturbed and stirred. We can speculate that
this induces turbulence that cascades on smaller scales and damps into particle
re-acceleration and fast magetic reconnection, which are two interconnected
processes if incompressible turbulence is considered (Brunetti \& Lazarian
\cite{bl16}).
 
The low luminosity of the ridge and of the  2 Mpc scale emission, and the
moderately steeper spectrum of the ridge, suggest that the
energy budget that becomes available to the non-thermal components in H2 is
smaller than that of classical giant radio halos, that indeed are generated
during major merger events.
An alternative possibility is that the radio emission is very old and marks
the switched off phase of the radio halo, when merger turbulence is
dissipated at later times.
Both cases are possible for massive clusters that show intermediate
properties between merging and relaxed systems (e.g. Cassano et al.
\cite{cassano06}, Brunetti et al. \cite{brunetti09}, Donnert et al.
\cite{donnert13}). This is indeed the case of A\,2142.
As a matter of fact, this radio halo is considerably underluminous compared
to the correlation between radio power and cluster mass
(Cassano et al. \cite{cassano13}), even considering the GBT measurement in
Farnsworth et al. (\cite{farnsworth13}) (see Sects. 4.2 and 6).

Different scenarios can explain the origin of the perturbations induced in the
ICM. A possibility is that the three cold fronts detected in X-rays
(see their location in Fig. 8)
are generated by a single event (Rossetti et al. \cite{rossetti13}).
In this case, H1 and H2 may probe the evolution of this
phenomenon on different scales and/or times and may trace different levels of
perturbations and magnetic field strength that are present in the ICM.
The symmetry of the three cold fronts and the way they encompass H1 and H2
is in support of this scenario.

Another possibility is that the ridge traces a turbulent region that is
generated by the continuous accretion of subhalos
along the S--E direction. This hypothesis may be supported by the fact that
optical data show the presence of several groups of galaxies that trace a
large-scale filament in the S--E direction (Owers et al. \cite{owers11},
Einasto et al. \cite{einasto15}), and group accretion in A\,2142 is
caught in action (Eckert et al. \cite{eckert14}).

Relativistic hadrons and their secondaries generated via inelastic collisions
in the ICM may also contribute to the origin of H1 and H2.
Following the scenario proposed for radio halos by Zandanel et al.
(\cite{zandanel14}),
the halo in A\,2142 may experience a transition from a hadronic emission
component, dominating the emission in the core region (H1), to a mainly
leptonic (re-acceleration) component, responsible for H2 and the large-scale emission. 
This scenario would qualitatively match the differences in the observed properties
of H1 and H2. Inefficient re-acceleration in the ridge, as explained above,
would produce a low-luminosity signal with steeper spectrum, whereas secondaries
generated in the dense core would explain the higher brightness of H1 and
its flatter spectrum.

It is interesting to note that the two-component radio halo 
in A\,2142 is not unique. A similar situation has been found for the 
diffuse emission in A\,2319 (Storm et al. \cite{storm15}).

\section{Summary and conclusions}

Our radio observations of A\,2142 show that the cluster is extremely 
interesting at radio wavelengths. 

We confirm that the giant radio halo in the cluster extends $\sim$1 Mpc 
in the NW--SE direction out to the region where the most distant cold 
front is located.
It is one of the few cases of a giant radio halo in a cluster that is not
a major merger (Bonafede et al. \cite{bonafede14}, Sommer et al.
\cite{sommer17}), thus it has the potential to provide important
constraints on the origin of these sources.

The overall properties show that the radio halo consists of two different
components. The first component is a bright and fairly compact central region
(H1) that is coincident
with the brightest part of the  X-ray, which is confined by the two inner cold
fronts and covers the volume where the two BCGs are also located; the second
component is a broader emission with lower surface brightness,  which we
named the ridge (H2) and extends out to the most distant cold front. 
The very large-scale emission imaged by the GBT at 1410 MHz Farnsworth et
al. \cite{farnsworth13} is undetected in all our interferometric images.
The spectral properties of these two components, derived from 118 MHz  to
1.78 GHz, show some differences.
The region H1 has a steep spectrum with
$\alpha_{234~MHz}^{1.478GHz}$=1.33$\pm$0.08, 
while the spectrum of H2 is moderately steeper, i.e.
$\alpha_{118~MHz}^{1.78~GHz}\sim1.5$, if we account for the relevance
of the u--v coverage of our set of observations in recovering the extent and
flux density of this component.

The radio halo in A\,2142 is significantly underluminous if compared to
classical radio halos hosted in clusters with similar mass: it is almost
a factor of 20 below the correlation in Cassano et al. \cite{cassano13}
if we consider the 1.4 GHz radio power of H1 and H2 presented in this work.
Even using the GBT measurement, which should be considered an upper limit,
we obtain logP$_{1.4~GHz}=1.15\times10^{24}$W~Hz$^{-1}$, i.e. a factor of
$\sim$4--5 lower than the correlation at the cluster mass,
M$\sim8.8\times10^{14}$
M$_{\odot}$ (see Fig. 7 in Bernardi et al. \cite{bernardi16} for an updated
version of the plot).

We suggest that two different mechanisms could be at the origin of the complex
diffuse radio emission in A\,2142.
We propose that on the core scale (H1) the emission is powered by
mechanisms that are similar to those considered for radio mini-halos. On
larger scales (H2), however, the emission is powered by mechanisms 
that are similar to those considered for classical giant radio halos,
but less powerful given that the dynamical properties of A\,2142
are indeed less extreme than those of major merging systems.
In this case turbulence may be generated by stirring of gas and magnetic fields
on the large scale or by off-axis minor mergers.
Alternatively, the properties of the radio halo may be interpreted
as due to the transition between a hadronic (secondary) component
in the core and a leptonic (turbulent re-acceleration) component
on the larger scale of the ridge. 

The phenomenology observed both in the X-rays and in the radio 
suggests that A\,2142 is a suitable target to understand, at the same time, 
the origin and evolution of cold fronts on different scales and the 
connection between radio halos and mini-halos.

Our work confirms that the study of the origin of cluster-scale 
radio sources is crucial to improve our understanding of the complex 
phenomena at play during the processes of cluster mergers and group
accretion in the Universe.
\\
\\

{\it Acknowledgements.}
TV and GB acknowledge partial support from PRIN--INAF 2014. 
LR and DF acknowledge partial support from the US National Science
Foundation under grant AST-1211595 to the University of Minnesota.
TS acknowledges support from the ERC Advanced Investigator programme
NewClusters 321271.
Basic research in radio astronomy at the Naval Research Laboratory is
supported by 6.1 base funding.
We thank the staff of the GMRT who have made these observations possible. 
GMRT is run by the National Centre for Radio Astrophysics of the Tata 
Institute of Fundamental Research. The National Radio Astronomy
Observatory is a facility of the National Science Foundation operated under 
cooperative agreement by Associated Universities, Inc. 
The Very Large Array is operated by the National Radio Astronomy
Observatory, which is a facility of the National Science Foundation
operated under cooperative agreement by Associated Universities, Inc.
This research has made use of the NASA/IPAC Extragalactic Database (NED) 
which is operated by the Jet Propulsion Laboratory, California Institute 
of Technology, under contract with the National Aeronautics and Space
Administration.
LOFAR, the Low Frequency Array designed and constructed by ASTRON, has
facilities in several countries, which are owned by various parties
(each with their own funding sources), and that are collectively operated
by the International LOFAR Telescope (ILT) foundation under a joint
scientific policy.
This research used the facilities of the Canadian Astronomy Data Centre 
operated by the National Research Council of Canada with the support of the 
Canadian Space Agency.


\begin{appendix} 
\onecolumn

\section{Optical identifications}
In this Appendix we report the full list of optical counterparts belonging
to A\,2142 in the same field shown in Fig. \ref{fig:fig2}, and radio/optical
overlays both for the extended radio galaxies and for those groups where
multiple radio emission has been detected.
%
%

\begin{longtable}{lllcccc}
\caption{\label{optid} Optical identifications with cluster galaxies. Errors
  on the radio flux density are of the order of 4--5\% at 608 MHz, and of
  the order of 10\% at 234 MHz.
  The optical information has been taken from the NASA Extragalactic 
Database (NED). Notes are
as follows: P=point-like source, NAT=narrow-angle tail, WAT=wide-angle tail.
$^{\star}$ Position of optical identification; $^{\diamond}$ Position 
of peak flux density.
The sequence number for each source is used in Fig. A.2 to show the
position on the optical plate.}\\
\hline\noalign{\smallskip}
\#, GMRT name  &  $\alpha_{\rm J2000}$ radio & $\delta_{\rm J2000}$ radio & S$_{\rm 608~MHz}$ (mJy) & S$_{\rm 234~MHz}$ (mJy) & logP$_{\rm 608~MHz}$(W/Hz) & Notes \\
\noalign{\smallskip}
Optical Catalogue & $\alpha_{\rm J2000}$ opt & $\delta_{\rm J2000}$ opt & m$_g$ & 
& z \\
\hline
\noalign{\smallskip}
\endfirsthead
\caption{continued.}\\
\hline\noalign{\smallskip}
\#, GMRT name  & $\alpha_{\rm J2000}$ radio & $\delta_{\rm J2000}$ radio & S$_{\rm 608~MHz}$ (mJy) & S$_{\rm 234~MHz}$ (mJy) & logP$_{\rm 608~MHz}$(W/Hz) & Notes \\
\noalign{\smallskip}
Optical Catalogue & $\alpha_{\rm J2000}$ opt & $\delta_{\rm J2000}$ opt & m$_g$ & 
& z \\
\hline
\noalign{\smallskip}
\endhead
\hline
\endfoot
\endlastfoot
\#1, GMRT-J\,155636+270041  & 15 56 36.99 & 27 00 41.7 & 1.67 &  6.52 & 22.55 & P \\
WISEPC      & 15 56 37.07 & 27 00 39.7 & 15.9 &       & 0.091117 \\
\\
\#2, GMRT-J\,155642+273324  & 15 56 42.94 & 27 33 24.1 & 3.18 &  5.71 & 22.80 & P \\
2MASX       & 15 56 42.96 & 27 33 24.6 & 17.0 &       & 0.088777 \\
\\
\#3, GMRT-J\,155646+270015  & 15 56 46.20 & 27 00 15.2 & 0.53 &   -   & 21.99 & P \\
SDSS        & 15 56 46.25 & 27 00 15.4 & 18.9 &       & 0.085227 \\
\\
\#4, GMRT-J\,155700+273102  & 15 57 00.56 & 27 31 02.2 & 1.87 &  4.42 & 22.50 & P \\
2MASX       & 15 57 00.56 & 27 31 02.7 & 17.8 &       & 0.082248 \\
\\
\#5, GMRT-J\,155703+271812  & 15 57 03.29 & 27 18 12.8 & 1.53 &  3.65 & 22.54 & P \\
2MASX       & 15 57 03.34 & 27 18 12.7 & 16.8 &       & 0.094222 \\
\\
\#6, GMRT-J\,155708+273519  & 15 57 08.97 & 27 35 19.0 & 2.92 &  6.58 & 22.79 & P \\
SDSS        & 15 57 08.92 & 27 35 19.3 & 17.6 &       & 0.123953\\
\\
\#7, GMRT-J\,155709+2702434  & 15 57 09.30 & 27 02 43.0 & 0.75 &  1.88 & 22.31 & P \\
SDSS        & 15 57 09.43 & 27 02 46.6 & 18.6 &       & 0.102941 \\
\\
\#8, GMRT-J\,155714+272608  & 15 57 14.12 & 27 16 08.6 & 1.68 &  3.29 & 22.60 & P \\
SDSS        & 15 57 14.14 & 27 16 07.9 & 18.4 &       & 0.095882 \\
\\
\#9, GMRT-J\,155722+270111  & 15 57 22.11 & 27 01 11.8 & 0.57 &  2.74 & 22.19 & P \\
SDSS        & 15 57 22.00 & 27 01 13.0 & 18.3 &       & 0.103162 \\
\\
\#10, GMRT-J\,155739+270655  & 15 57 39.93 & 27 06 55.7 & 0.60 &   -   & 22.14 & P \\
2MASX       & 15 57 39.24 & 27 06 56.7 & 17.5 &       & 0.095269 \\
\\
\#11, GMRT-J\,155739+272249  & 15 57 39.84 & 27 22 49.7 & 0.34 &   -   & 21.86 & P \\
2MASX       & 15 57 39.91 & 27 22 48.9 & 17.4 &       & 0.091306 \\
\\ 
\#12, GMRT-J\,155744+270121  & 15 57 44.90 & 27 01 21.3 & 0.52 &   -   & 22.10 & P \\
SDSS        & 15 57 44.61 & 27 01 19.0 & 22.2 &       & 0.097427 \\
\\
\#13, GMRT-J\,155745+274042  & 15 57 45.95 & 27 40 42.3 & 1.59 &  5.36 & 22.51 & P \\
WISPEC      & 15 57 45.93 & 27 40 43.6 & 17.3 &       & 0.089887 \\
\\
\#14, GMRT-J\,155747+271857  & 15 57 47.07 & 27 18 57.3 & 0.88 &  1.71 & 22.36 & P \\
2MASX       & 15 57 46.93 & 27 18 56.6 & 17.4 &       &0.100987 \\
\\
\#15, GMRT-J\,155801+271500  & 15 58 01.15 & 27 15 00.4 & 1.10 &  2.05 & 22.31 & P \\
2MASX       & 15 58 01.28 & 27 15 00.8 & 17.8 &       & 0.085613 \\
\\
\#16, GMRT-J\,155807+271112  & 15 58 07.90 & 27 11 12.5 & 1.75 &  3.44 & 22.68 & P \\
SDSS        & 15 58 07.93 & 27 11 12.1 & 17.8 &       & 0.102417 \\
\\
\#17, GMRT-J\,155812+271536  & 15 58 12.39 & 27 15 36.5 & 0.34 &   -   & 21.85 & P \\
SDSS        & 15 58 12.23 & 27 15 37.3 & 19.9 &       & 0.090876 \\
\\
\#18, GMRT-J\,155814+271619  & 15 58 14.31 & 27 16 19.0 & 272.71 & 550.11 & 24.80 & NAT$^{\star}$ - T1 \\
SDSS        & 15 58 14.31 & 27 16 19.0 & 17.5   &       & 0.095546 \\
\\
\#19, GMRT-J\,155816+271412  & 15 58 16.56 & 27 14 12.5 & 0.83 &  1.91 & 22.22 & P \\
SDSS        & 15 58 16.58 & 27 14 11.5 & 18.1 &       & 0.089338 \\
\\
\#20, GMRT-J,155818+271421  & 15 58 18.69 & 27 14 21.5 & 0.47 &   -   & 21.88 & P \\
2MASX       & 15 58 18.75 & 27 14 21.0 & 17.8 &       & 0.080353 \\
\\
\#21, GMRT-J\,155819+271400  & 15 58 19.93 & 27 14 00.5 & 1.00 &   -   & 22.32 & P \\
2MASX       & 15 58 20.02 & 27 14 00.0 & 16.2 &       & 0.090369 \\
\\
\#22, GMRT-J\,155820+272000  & 15 58 20.84 & 27 20 00.7 & 158.64 & 501.47 & 24.51 & Tail$^{\star}$ - T2 \\
2MASX       & 15 58 20.91 & 27 20 01.0 & 16.6   &       & 0.089553 \\
\\
\#23, GMRT-J\,155824+271127  & 15 58 24.20 & 27 11 27.5 & 2.60 &  5.05 & 22.74 & Ext.$^{\star}$ \\
SDSS        & 15 58 24.50 & 27 11 25.5 & 18.0 &       & 0.091530 \\
\\
\#24, GMRT-J\,155829+271713  & 15 58 29.38 & 27 17 13.9 & 0.90 &  1.46 & 22.27 & P \\
SDSS        & 15 58 29.36 & 27 17 14.2 & 17.1 &       & 0.089875 \\
\\
\#25, GMRT-J\,155829+270654  & 15 58 29.70 & 27 06 54.4 & 0.92 &  0.96 & 22.22 & P \\
2MASX       & 15 58 29.55 & 27 06 55.0 & 17.8 &       & 0.084849 \\
\\
\#26, GMRT-J\,155831++265633  & 15 58 31.70 & 26 56 33.4 & 0.77 &  1.99 & 22.15 & P \\
2MASX       & 15 58 31.73 & 26 56 32.1 & 17.3 &       & 0.085579 \\
\\
\#27, GMRT-J\,155837+270818  & 15 58 37.01 & 27 08 18.4 & 0.53 &  2.01 & 22.04 & P \\
2MASX       & 15 58 37.04 & 27 08 17.1 & 17.3 &       & 0.090295 \\
\\
\#28, GMRT-J,155839+271618  & 15 58 39.17 & 27 16 18.4 & 0.38 &   -   & 21.88 & P \\
2MASX       & 15 58 39.29 & 27 16 18.1 & 17.6 &       & 0.089194 \\
\\
\#29, GMRT-J\,155842+270736  & 15 58 42.85 & 27 07 36.3 & 3.73 &  8.86 & 22.85 & P \\
WISEPC      & 15 58 42.84 & 27 07 36.5 & 17.4 &       & 0.086434 \\
\\
\#30, GMRT-J\,155843+272254  & 15 58 43.59 & 27 22 54.3 & 0.39 &   -   & 21.90 & P \\
SDSS        & 15 58 43.57 & 27 22 53.2 & 18.7 &       & 0.089284 \\
\\
\#31, GMRT-J\,155843+270752  & 15 58 43.75 & 27 07 52.8 & 2.23 &  4.30 & 22.63 & P \\
SDSS        & 15 58 43.70 & 27 07 52.8 & -    &       & 0.085881 \\
\\
\#32, GMRT-J\,155844+270812  & 15 58 44.87 & 27 08 12.3 & 1.65 &  3.42 & 22.58 & P \\
WISEPC      & 15 58 44.97 & 27 08 12.4 & 17.7 &       & 0.094569 \\
\\
\#33, GMRT-J\,155844+271831  & 15 58 44.13 & 27 18 31.8 & 0.89 &  2.49 & 22.15 & P \\
SDSS        & 15 58 44.15 & 27 18 31.7 & 18.3 &       & 0.079943 \\
\\
\#34, GMRT-J\,155845+263851  & 15 58 45.42 & 26 38 51.3 & 25.16 & 24.30 & 23.81 & P \\
2MASX       & 15 58 45.46 & 26 38 51.0 & 16.3  &       & 0.1000 \\ 
\\
\#35, GMRT-J\,155846+271552  & 15 58 46.93 & 27 15 52.8 & 1.20 &  2.95 & 22.30 & P \\
SDSS        & 15 58 46.96 & 27 15 52.8 & 18.0 &       & 0.081561 \\
\\
\#36, GMRT-J\,155849+273639  & 15 58 49.69 & 27 26 39.2 & 2.06 &  4.16 & 22.57 & P \\
WISEPC      & 15 58 49.65 & 27 26 40.3 & 17.0 &       & 0.084872 \\
\\
\#37, GMRT-J\,155850+273324  & 15 58 50.46 & 27 23 24.2 & 31.89 & 53.89 & 23.85 & Ext$^{\diamond}$ - G \\
WISEPC      & 15 58 50.39 & 27 23 24.5 & 16.1  &       & 0.093601 \\
\\
\#38, GMRT-J\,155851+272349  & 15 58 51.92 & 27 23 49.7 & 3.25 &  6.37 & 22.91 & P \\
2MASX       & 15 58 51.89 & 27 23 50.3 & 17.2 &       & 0.098645 \\
\\
\#39, GMRT-J\,155855+272124  & 15 58 55.85 & 27 21 24.1 & 0.41 &   -   & 21.98 & P \\
            & 15 58 55.7  & 27 21 23   & 18.9 &       & 0.0952906 \\
\#40, GMRT-J\,155903+272140  & 15 59 03.85 & 27 21 40.5 & 0.33 &  0.98 & 21.83 & P \\
SDSS        & 15 59 03.74 & 27 21 37.7 & 19.2 &       & 0.089533 \\
\\
\#41, GMRT-J\,155920+271138  & 15 59 20.75 & 27 11 38.6 & 0.66 &  1.25 & 22.15 & P \\ 
GALEX       & 15 59 20.73 & 27 11 40.2 &  -   &       & 0.092125 \\
\\
\#42, GMRT-J\,155930+270530    & 15 59 30.92 & 27 05 30.7 & 0.76 &  1.49 & 22.18 & P \\
SDSS        & 15 59 30.98 & 27 05 29.1 & 18.4 &       & 0.088270 \\
\\
\hline
\end{longtable}
\twocolumn

\vfill\eject
\bigskip
\bigskip
\begin{figure*}
\centering
\includegraphics[angle=0,height=18cm]{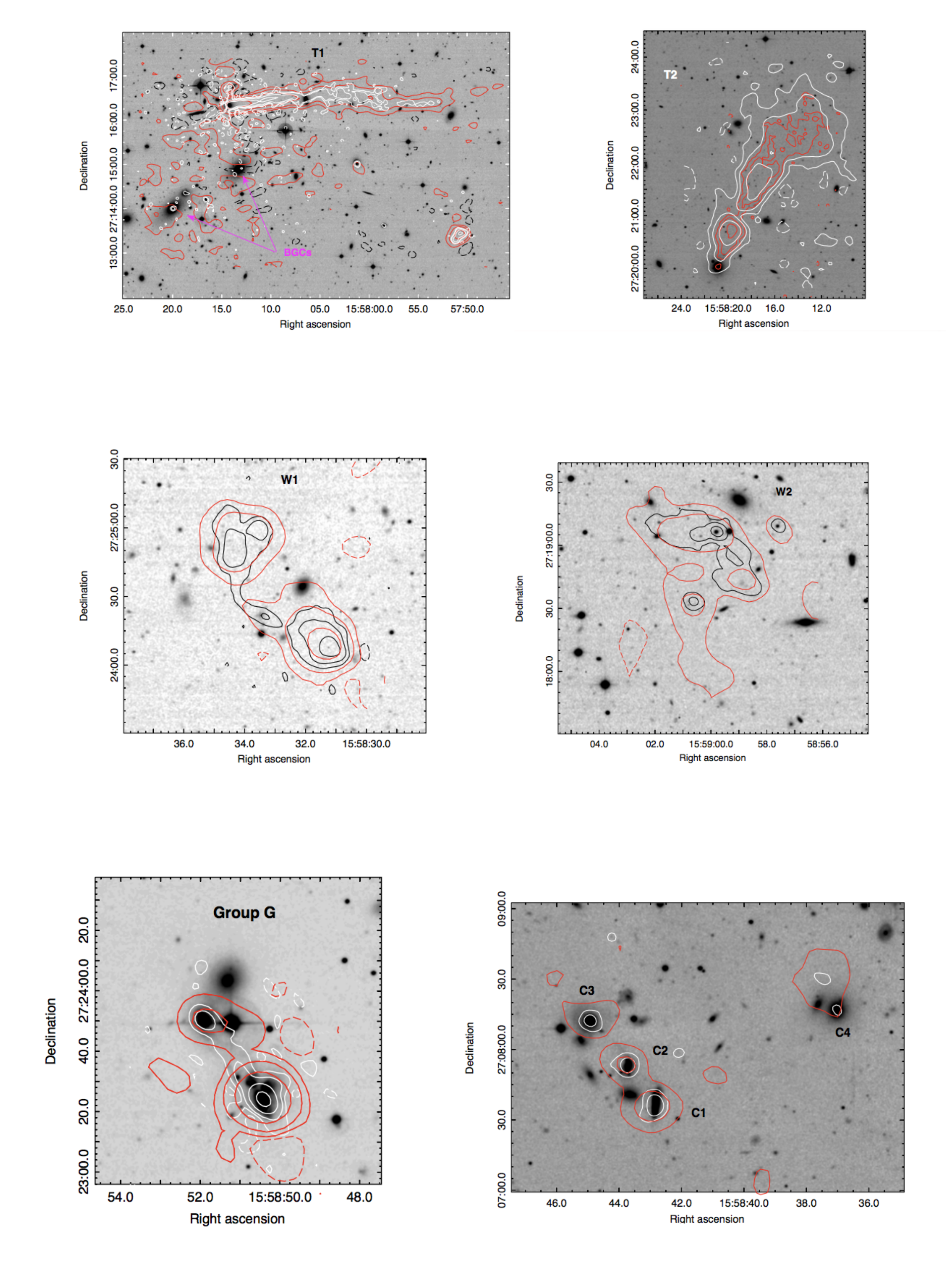}
\caption{Overlays between a CFHT MegaCam g-band image and the
  radio emission at 608 MHz (resolution of
  $5.2^{\prime\prime}\times4.5^{\prime\prime}$, p.a. 52.6$^{\circ}$)
  and at 234 MHz (resolution of
  $12.8^{\prime\prime}\times11.6^{\prime\prime}$, p.a. 67.3$^{\circ}$).
  In all panels the 608 MHz contours are spaced by a factor of 4
  starting from $\pm$0.15 mJy~b$^{-1}$ and the 234 MHz contours
  are spaced by a factor of 4 starting from $\pm$0.75 mJy~b$^{-1}$.
  Panels are as follows: {\it Upper left} - T1 (B2~1556+27), 608
  MHz contours are indicated in white (negative dashed), 234 MHz contours are
  indicated in red (negative dashed black). The two BCGs are shown.
  {\it Upper right} - T2 (B2~1556+27), 608 MHz contours are shown in red
  (negative dashed), 234 MHz contours are shown in white (negative dashed).
  {\it Central left} - W1, 608 MHz contours are shown in black (negative
  dashed), 234 MHz contours are indicated in red (negative dashed black). 
  {\it Central right} - W2, 608 MHz contours are indicated in black (negative
  dashed), 234 MHz contours are indicated in red (negative dashed).
  {\it Bottom left} - Group  G, 608 MHz contours are shown in white (negative
  dashed), 234 MHz contours are shown in red (negative dashed black). 
  {\it Bottom right} - Group C, 608 MHz contours are indicated in white (negative
  dashed), 234 MHz contours are shown in red (negative dashed).}
\label{fig:fig1app}
\end{figure*}
%
%
\begin{figure*}
  \includegraphics[angle=0,height=15cm]{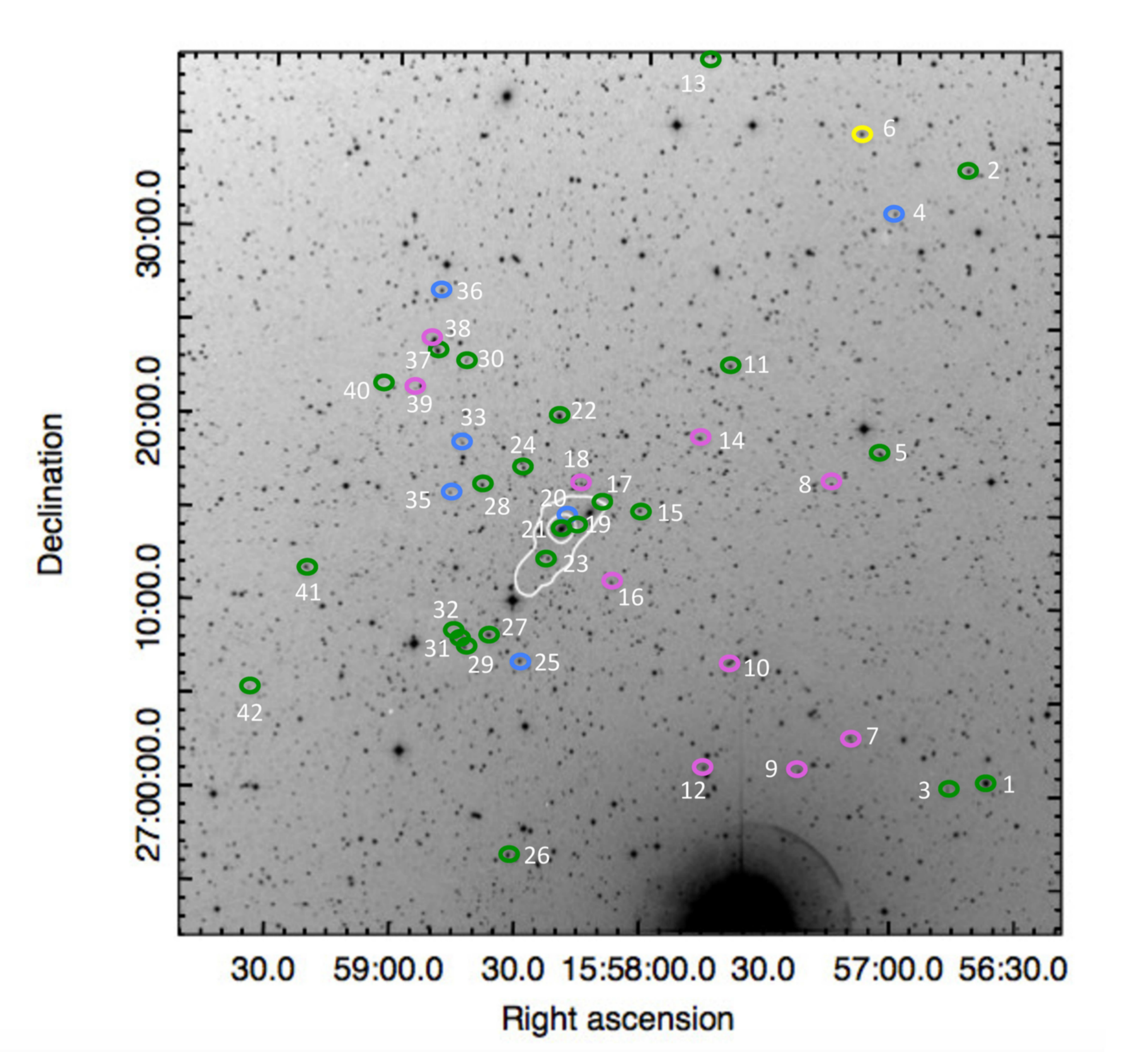}
\caption{Location of the radio galaxies in A\,2142 listed in Table A.1 
  on the DSS2 red plate (circles). The numbering follows that of
  the table. The 1377 MHz radio contours of the halo are shown in white
  (0.8 and 1.6 mJy~b$^{-1}$, resolution 60$^{\prime\prime}\times60^{\prime\prime}$). 
  Different redshift intervals are colour coded as follows:
  blue=0.075 - 0.085; green=0.085 - 0.095; magenta=0.095 - 0.105;
and  yellow=0.105 - 0.125.}
\label{fig:fig7a}
\end{figure*}

\end{appendix}

\end{document}